\documentclass[fleqn, usenatbib]{mnras}
\usepackage{siunitx}
\usepackage{graphicx}	% Including figure files
\usepackage{amsmath}	% Advanced maths commands
\usepackage{amssymb}	% Extra maths symbols
\usepackage[version=3]{mhchem}
\usepackage{newtxtext}
\usepackage[slantedGreek]{newtxmath}
\usepackage{multirow}
\usepackage{subfig}
\usepackage{tabularx}
\usepackage{xcolor}

\newcolumntype{Y}{>{\centering\arraybackslash}X}

\newcommand{\er}[3]{\ensuremath{#1^{+#2}_{-#3}}}
\newcommand{\Msun}{\ensuremath{\,{\rm M}_\odot}}                  % Solar mass symbol
\newcommand{\Mjup}{\ensuremath{\,{\rm M}_{\rm Jup}}}              % Jupiter mass symbol
\newcommand{\Rjup}{\ensuremath{\,{\rm R}_{\rm Jup}}}              % Jupiter radius symbol
\newcommand{\pjup}{\ensuremath{\,\rho_{\rm Jup}}}                 % Jupiter density symbol
\newcommand{\mc}[1]{\multicolumn{2}{c}{#1}}

\graphicspath{{img/}}

\author[Morrell et al.]{
Sam Morrell${^
{\href{https://orcid.org/0000-0001-6352-5312}{\includegraphics[scale=0.5]{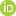}}}}$,
$^{1}$\thanks{E-mail: \href{mailto:smorrell@astro.ex.ac.uk}
{smorrell@astro.ex.ac.uk}}
Tim Naylor$^
{\href{https://orcid.org/0000-0002-0506-8501}{\includegraphics[scale=0.5]{orcid.jpg}}}$,$^{1}$
John Southworth$^
{\href{https://orcid.org/0000-0002-3807-3198}{\includegraphics[scale=0.5]{orcid.jpg}}}$$^{2}$
and 
David K. Sing$^
{\href{https://orcid.org/0000-0001-6050-7645}{\includegraphics[scale=0.5]{orcid.jpg}}}$$^{3}$ \\
$^{1}$Department of Physics and Astronomy, University of Exeter, Exeter, EX4 4QL, UK \\
$^{2}$Astrophysics Group, Keele University, Staffordshire, ST5 5BG, UK \\
$^{3}$Department of Physics \& Astronomy, Bloomberg Center for Physics and Astronomy, 3400 N. Charles Street, Baltimore, MD 21218, USA
}

\date{Accepted 2025 November 7. Received 2025 September 30; in original form 2024 September 6}

\pubyear{2025}

\begin{document}
\label{firstpage}
\pagerange{\pageref{firstpage}--\pageref{lastpage}}
\title[Stellar SEDs and Exoplanet Parameters]{Using Stellar Spectral Energy Distributions to Measure Exoplanet Parameters}
\pagerange{\pageref{firstpage}--\pageref{lastpage}}
\maketitle

% For Simon Dawson, who revelled in the universe, and the wonders of space and time. 
\begin{abstract}
The ability to make accurate determinations of planetary parameters is inextricably linked to measuring physical parameters of the host star, in particular the stellar radius.
In this paper we fit the stellar spectral energy distributions of exoplanet hosts to measure their radii, making use of only archival photometry, the $Gaia$ parallaxes and $Gaia$ extinction maps.
Using the extinction maps frees us of the degeneracy between temperature and extinction which has plagued this method in the past. 
The resulting radii have typical random uncertainties of about 2 per cent.
We perform a quantitative study of systematic uncertainties affecting the methodology and find they are similar to, or smaller than, the random ones.
We discuss how the stellar parameters can be used to derive the properties of both transiting exoplanets, and those where only a radial-velocity curve is available.
We then explore in detail the improvements the method makes possible for the parameters of the PanCET sample of transiting planets.
For this sample we find the best literature measurements of the planetary radii have mean uncertainties about 40 per cent larger than those presented here, with the new measurements achieving  precisions of 2 per cent in radius and 10 per cent in mass.
In contrast to much recent work, these transiting exoplanets parameters are derived without using theoretical models of stellar interiors, freeing them of the assumptions those models contain, and any priors for stellar age.   
As the data used are available for the whole sky, the method can be used for self-consistent measurements of the planetary parameters of a very large fraction of known exoplanets.
\end{abstract}

% TODO: Update keywords for new paper. 

\section{Introduction} 

It is well known that determinations of the masses and radii of exoplanets rely on knowing the masses and radii of the stars that host them.
There are three key ways that this can affect our sample of exoplanet parameters.
First the simple precision with which we can measure these stellar parameters, in the absence of any systematic errors, will directly drive the uncertainties in exoplanet parameters.
The second issue is the systematic errors themselves.
Does the use of differing assumptions mean our exoplanet parameters are not strictly comparable, and if they are comparable, are they systematically shifted from the truth?
Finally there is the issue of the amount of telescope time required to characterise the stars.
Are there observationally cheaper ways of getting at least adequate parameters for the increasing number of exoplanet discoveries?

In this paper we argue that photometric techniques may now yield the most accurate stellar temperatures and radii.
Since the required data are available from all-sky archives this allows an homogeneous set of parameters to be derived for all exoplanets without the need for a high-resolution stellar spectrum (although the metallicity derived from it may improve the parameters further).
Finally, we shall show that for systems with transits and radial velocity curves it allows the determinations of the planetary parameters to be independent both of the models of stellar structure and priors on the stellar age.

To explore this method we first give an outline description of how we measure the stellar parameters in section \ref{sec:a_new_technique}, and how they could be used to measure the parameters of exoplanets in section \ref{sec:application}.
The transiting systems with radial velocity curves present the most complex problem for finding planetary parameters.  
Furthermore it is those systems for which we are proposing the most radical departure from current methods, in that we are arguing for removing some constraints from the parameter determination.  
We therefore chose a group of these systems to demonstrate the revised method on, in section \ref{sec:pancet}.  
Afterwards we address the effect of likely errors in the determination of the extinction (section \ref{sec:extinction-discussion}), and metallicity (section \ref{sec:metallicity-discussion}).  
We also consider the effects of stellar activity (section \ref{sec:activity-discussion}) and the possibility of adding UV photometry to our input data (section \ref{sec:uv-photometry-discussion}). 
We summarise our conclusions in section \ref{sec:conclusion}.

% Introduce the technique. 
\section{The Technique} % (fold)
\label{sec:a_new_technique}
In \citet{Morrell:2019aa,Morrell:2020aa} we presented a spectral energy distribution (SED) method, and used it to measure radii ($R_{\star}$) and SED temperature ($T_{\rm SED}$) of a sample of $>15\,000$ main-sequence (MS) M-dwarfs to a median random uncertainty of 1.9 per cent and 0.5 per cent, respectively.
In Section 4.2 of \cite{2025MNRAS.540.1786F} we described using the same method to derive $R_{\star}$ and $T_{\rm SED}$ for a sample of bright FGK main-sequence stars ({\tt Teff\_SED} and {\tt radius\_SED} in the associated online table).
The method uses multi-band photometry to measure the luminosity and temperature of stellar SEDs, and thus has its roots in the long history of fundamental parameter determination from such measurements (see Appendix \ref{sec:history}).
We will briefly describe it here---for a more detailed description, see \citet{Morrell:2019aa}. 

A given photometric band samples the portion of the stellar SED over which its system response falls, the flux of which is determined primarily by the overall shape of the underlying blackbody, but additionally also incorporates strong, temperature-dependent spectral features. 
By sampling the SED across many photometric bands simultaneously, we are able to build up a 'fingerprint' of the stellar SED, which, when compared to equivalent synthetic photometry from model stellar atmospheres, is able to precisely measure both the total flux from the star, and the shape of the stellar SED. 
This flux, combined with a precise geometric distance ($d$) from $Gaia$ \citep{BailerJones2021a}, yields the stellar bolometric luminosity ($L$).
The shape of the SED maps onto an SED temperature $T_{\rm SED}$; so-called to distinguish it from other measures of temperature. 
Measurements of both $L$ and $T_{\rm SED}$ yield a unique measurement of $R_\star$ from the definition of effective temperature $T_{\rm eff}$. 

\begin{figure*}
    \includegraphics[width=0.9\textwidth]{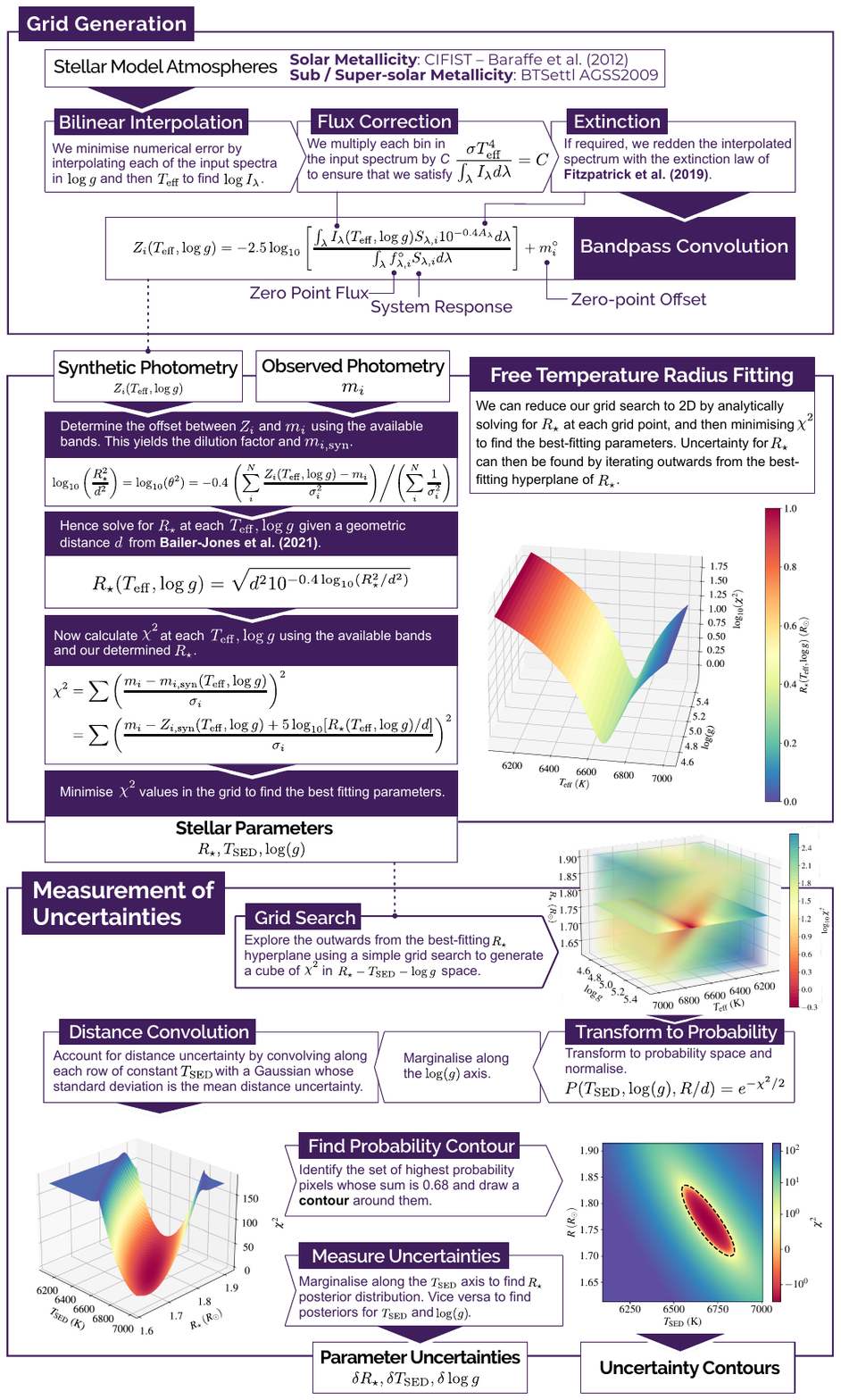}
    \caption{A schematic showing the process of the SED fitting method, starting with generating grids of synthetic photometry (top), how we measure stellar parameters (middle), and finally how we determine their uncertainties (bottom). }
    \label{fig:sedf-schematic}
\end{figure*}

The grid of synthetic photometry with which we perform the fitting process occupies the $T_{\rm eff} - \log(g)$ space, which we iterate through using a simple grid search algorithm. 
The models are supplied in units of mean disc intensity at the stellar surface $I_{\lambda}$, which incorporates the effect of limb darkening. 
By directly folding $I_{\lambda}$ through system responses, we can analytically determine the best-fitting radius by inferring the magnitude offset between the theoretical magnitudes and the observed photometry; which itself yields the dilution factor $R_\star^2 / d^2$. 
The parameters of the best fitting model SED, denoted by the lowest $\chi^2$ between its synthetic photometric and the observed photometry, are then adopted for the star. 
The radius uncertainty can then be found by iterating the grid search outwards from the analytically determined radius, producing a posterior in $T_{\rm SED} - \log(g) - R_\star$ space.
A schematic of the whole fitting process is shown in Fig.\,\ref{fig:sedf-schematic}. 

Until recently a problem with using $T_{\rm SED}$ would have been that the SED fit would have to include the interstellar extinction as a free parameter, and extinction and temperature are notoriously correlated \citep[e.g.][]{2006A&A...450..735M}.
The problem has been removed by the availability of 3D extinction maps, driven by the $Gaia$ parallaxes \citep[e.g.][]{2023A&A...674A..31D, 2019ApJ...887...93G, 2022AA...664A.174V}.
This means that given the $Gaia$ distance we have an independent measure of the extinction, which can be fixed in our fit to the SED.

\section {Application to exoplanet parameter determination}
\label{sec:application}

Having shown how we might obtain precise stellar temperatures and radii, we can now outline how we might utilise the Doppler reflex radial velocity curve of the host and/or the parameters of any observed transit to determine the planetary parameters.  

\subsection{Transit-only systems}

If there are no radial velocity or astrometric data for a system where a planet transits the star, the fundamental parameters that can be determined for the planet are its radius and the star-planet distance.
The transit depth yields the planetary radius as a fraction of the stellar radius, and its duration encodes the star-planet distance, again as a fraction of the stellar radius \citep[see, for example,][]{Mandel:2002aa}.
Hence determining the stellar radius is key to unlocking the parameters of the planet.
If we know the temperature and luminosity of the star, then the stellar radius will follow from the Stefan-Boltzmann Law.
An oft-used route is to use a spectroscopic temperature ($T_{\rm sp})$, perhaps along with a spectroscopically determined gravity and metallicity, to find an appropriate model atmosphere.
This is then folded through the band-passes of the available broad-band photometry, and the best fitting normalisation used along with a $Gaia$ parallax to determine the luminosity \citep[e.g.][] {2021AA...645A..16P, 2019AA...625A..68S, 2017AJ....153..136S}.
For stars with significant extinction that has to be left as a free parameter in the SED fit.

Where this route differs from that proposed here is that we use $T_{\rm SED}$ rather than a spectroscopic temperature determination.
This raises the question of whether  $T_{\rm SED}$ is better than a temperature derived from spectra, or at least whether it is good enough.
As we show in section \ref{sec:results}, our SED temperatures can achieve precisions of 50 K and 
accuracies much better than that when compared with the  \cite{2010A&A...512A..54C} temperature scale.
Hence we conclude that the SED method, especially if combined with spectroscopically determined metallicities and gravities when selecting the model atmospheres, should give better stellar radii.

We can also have the possibility of determining the mass of the star in such systems, provided the mass of the planet is much smaller than that of the star, and that its orbit is circular.
Given that we already know the period and the radius of the planet's orbit, Newton's Law will give us the mass of the star.

\subsection{Radial velocity only systems}

Where we have just the radial velocity curve for a system, we can determine $M_{\rm P}\sin i$ and $a_{\rm P} \sin i$ (where $M_{\rm P}$ is the mass of the planet, $a_{\rm p}$ is the semimajor axis of the planet's orbit and $i$ is the inclination of the angular momentum vector of the orbit to the line of sight).
These quantities can be obtained by measuring the mass function from the radial velocity curve, provided we have the mass of the star, $M_\star$.  
$M_\star$ can only come from stellar structure models, but in principle spectra can give  $T_{\rm eff}$ and $\log(g)$ from which a mass can be interpolated from the models, with the option of refining it using a spectroscopically determined metallicity.  
In practice $\log(g)$ cannot be determined from spectra to sufficient accuracy to be useful.  
Thus normal practice is to use a spectroscopic temperature and metallicity along with the $Gaia$ parallax to determine the luminosity in the way described above, and then determine $M_\star$ from isochrones \citep[e.g.][]{2024MNRAS.531.4464D}.
Once again we would argue that, given the possibility of correcting the SED for extinction using the $Gaia$-based extinction maps, it is better to use $T_{\rm SED}$.

\subsection{Transiting systems with RV curves}
\label{sec:RV+trans}

Systems with a radial velocity curve and a transit give us the opportunity to measure both the mass and radius of the planet.
If we use $T_{\rm SED}$ and the luminosity (from the SED and the $Gaia$ parallax) to measure the radius of the star, the parameters from the transit give us the star-planet distance and the planetary radius.  The transit also yields the inclination, which means the radial velocity curve gives us the semimajor axis of the star's barycentric orbit.
That yields the mass ratio and, from Newton's Law of Gravitation (via the mass function), the mass of both the planet and the star.

In fact this is not the route most modern measurements of planetary mass and radius follow.
They over-constrain the system using -- in addition to $T_{\rm SED}$ -- a spectroscopic temperature, the spectroscopic gravity and an additional constraint to match the stellar interior models.
They normally achieve this by working within a Bayesian framework \citep[e.g.][]{2024AA...682A.129M, 
2022A&A...663A.101P, 
2023AA...677A..33b}
which also includes a prior on the stellar age.
The question we ask in this paper for these systems is whether there is a minimal set of constraints which can determine the planetary parameters to similar accuracy to the full Bayesian treatment.
In particular would there be significant advantages to being free of the structure models? 
Is there also an advantage in having a single temperature scale on which the measurements are made?

\section{Application to the PanCET sample}
\label{sec:pancet}

\subsection{The Sample}
\label{sec:pancet_intro}

Having outlined the methods for determining the exoplanet parameters, we will now illustrate them in detail at the same time as testing them, by deriving parameters for a sample of exoplanets.
The most important criterion for selecting the stars in that sample is that they are transiting systems.
As explained in section \ref{sec:RV+trans} these are the systems where we are suggesting the largest departure from current practice.
It would also be of advantage if they are bright, since that implies high signal-to-noise, which is ideal for existing methods, and if they cover a range of stellar spectral types similar to the population of known hosts.
The \textit{Panchromatic Exoplanet Treasury} programme \citep[PanCET, ][]{PanCET} matches all these criteria.
Literature measurements indicate that the stars span $T_{\rm eff}$ values of 3\,000 to 7\,000\,\si{\kelvin} -- the domain occupied by around $90$ {per cent} of the known exoplanet hosts. 
While we have demonstrated the accuracy of the technique for characterising low-mass stars \citep{Morrell:2019aa}, this broad parameter space coverage allows us to assess how well the new methodology generalises to the population of known exoplanet hosts as a whole. 

\begin{figure}
	\includegraphics[width=\columnwidth]{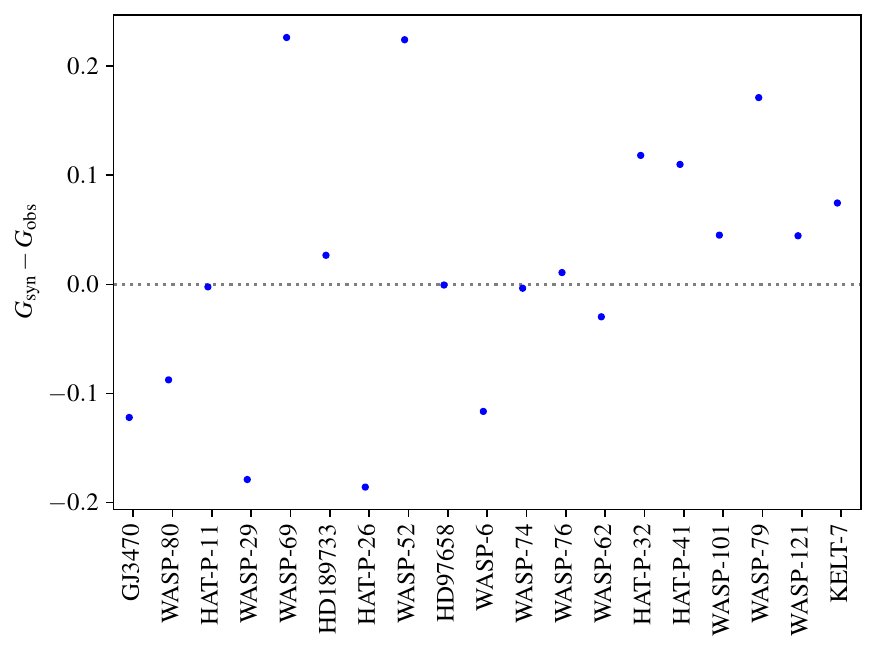}
	\caption[The residual between the synthetic and observed $G$ photometry of exoplanet hosts as a function of temperature. ]{The residual between the synthetic and observed $G$ photometry of our chosen exoplanet hosts as a function of temperature. The synthetic photometry was generated using stellar atmospheres from the $T_{\rm sp}$ and $R_\star$ presented in the literature and placed at the distance from $Gaia$. }
	\label{fig:lit-lum-comp-names-litonly}
\end{figure}

As a simple test of the veracity of the current stellar parameters for the PanCET sample, we calculated the $G$-band magnitudes predicted by the parameters given in the Transiting Extrasolar Planet Catalogue (TEPCat\footnote{\texttt{http://www.astro.keele.ac.uk/jkt/tepcat/}}; \citealt{Me11mn}), and compared them with the data.  
For each star in the sample we selected the CIFIST \citep{Allard:2012aa} model atmosphere at the appropriate literature $T_{\rm sp}$ and folded it through the \textit{Gaia} $G$-band filter.
We then applied the dilution factor appropriate for the $R_{\star}$ from the literature, at the distance as prescribed by \citet{BailerJones2021a}, to yield a synthetic magnitude measurement. 
We then compared each synthetic magnitude directly to the equivalent observed $G$-band photometry from $Gaia$ DR3 in Fig.\,\ref{fig:lit-lum-comp-names-litonly}.
A third of the sample lie further than $1 \sigma$ from their observed $G$-band magnitude; which in itself should not be surprising, given that the presented uncertainties should encompass two thirds of the probability in the luminosity posterior. 
However, our main concern is the RMS of these residuals, which is on the order of $20$ {per cent} in luminosity. 
In \citet{Morrell:2019aa}, we showed that for M-dwarfs, whose measurements are historically less reliable than those of Solar-type stars, we were able to achieve luminosity measurements to a precision of $\sim 1 - 2$ per cent; showing the potential for an order of magnitude improvement over the literature measurements. 

\subsection{Stellar Parameters}
\label{sec:star_params}

\subsubsection{Input Catalogue}
\label{sec:input-catalogue}

\begin{figure}
 	\includegraphics[width=\columnwidth]{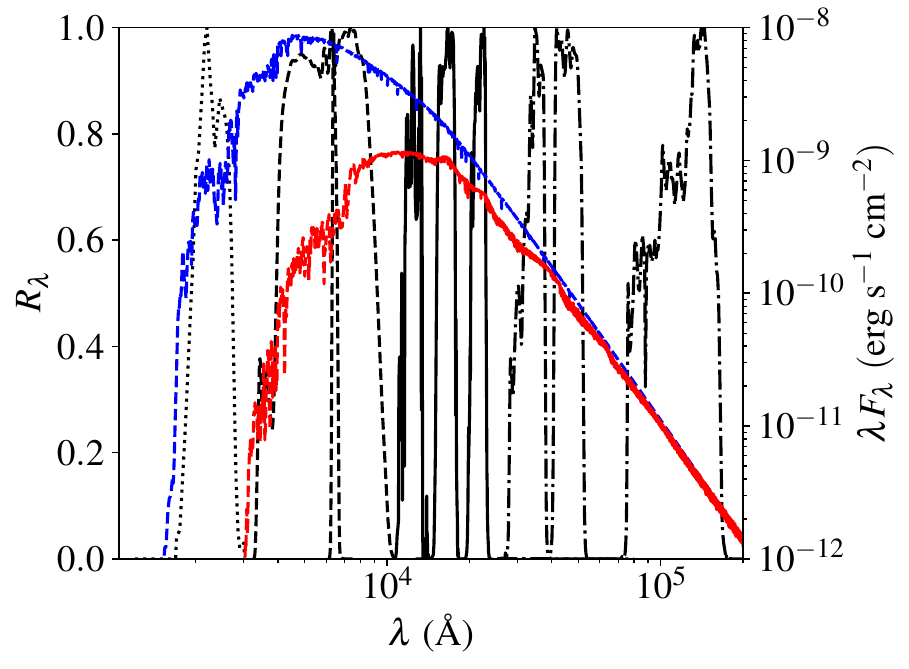}
 	\caption{The coverage of all of the system responses used in the fitting process. 
	The chosen systems sample the UV (GALEX, dotted), optical ($Gaia$ DR2, dashed), near-IR (2MASS, solid) and mid-IR (WISE, dot-dashed). 
	The model spectra correspond to the best fitting models for GJ 3470 (red) and KELT-7 (blue), which lie at the $T_{\rm SED}$ extrema of the sample. They illustrate that the bands used in the fit are more than adequate to sample the SED of the stars in our input catalogue. }
 	\label{fig:filter-coverage}
\end{figure}

The input catalogue is required to provide the fitting process with multi-band photometric measurements, to adequately sample the shape and flux of the stellar SED, and precise distances, provided by the geometric distances of \citet{BailerJones2021a}. 
The wide $T_{\rm eff}$ range of our targets motivates the use of ultraviolet (UV) to mid-infrared (IR) photometry to thoroughly sample the SED of all targets; our justification for which is demonstrated in Fig.\,\ref{fig:filter-coverage}.
We chose those surveys with robust all-sky photometry, GALEX, $Gaia$, 2MASS and WISE (although we omitted W4 as it has low signal-to-noise ratio).
The surveys from which we drew photometry, along with the selection criteria for the data, are detailed in \autoref{tab:ch3-data-sources}. 
%!TEX root = ../exoplanet-radii.tex
\begin{table*}
	\centering
 	\caption[The sources of and selection criteria for the photometric data used in \autoref{ch:chapter3}]{The sources of the photometric data used for the SED fitting. To aid in comparison to plots, each filter is listed with its isophotal effective wavelength $\lambda_\text{iso}$, criteria used to select the photometry, and original source. The bands are grouped into photometric systems and in order of ascending $\lambda_\text{iso}$. The $\lambda_\text{iso}$ is determined by integrating $f_\lambda = 1$ across the filter. \\
    * We only include the NUV band in SED fitting performed in \autoref{sec:uv-photometry-discussion}, for reasons discussed in that section. }
	\begin{tabularx}{\linewidth}{ Y  Y  Y  Y  >{\hsize=1.25\hsize}Y }
	\hline
	Band & $\lambda_\text{iso}$ (\AA) & System & Selection Criteria & Source \\
	\hline
	%$g_\text{PS1}$ & 4810 & \multirow{4}{*} { PS1 } & \multirow{4}{*} {
	%\begin{tabular}{c}
	%	$\textrm{nStackDetections} > 3$\\
	%\end{tabular}
	%} & \multirow{4}{*} { \citet{Chambers:2018aa} } \\
	%$r_\text{PS1}$ & 6170 & & & \\
	%$i_\text{PS1}$ &  7520 & & & \\
	%$z_\text{PS1}$ &  8660 & & & \\

	\multirow{1}{*}{NUV*} & \multirow{1}{*}{ 2316 } & \multirow{1}{*}{ GALEX } & \multirow{1}{*}{ $\delta {\rm NUV} < 0.1$ } & 
	\begin{tabular}{l}
		Passbands: \citet{Morrissey:2007aa} \\ \citet{Bianchi:2017aa}	
	\end{tabular}
	\\

	\hline
	\multirow{3}{*} { \begin{tabular}{l} $G_\mathrm{BP}$ \\ $G_\mathrm{RP}$ \end{tabular} } & 
	\multirow{3}{*} { \begin{tabular}{c} $5183$ \\ $7825$ \end{tabular} }  & 
	\multirow{3}{*} { Gaia DR3 } & \multirow{3}{*} {
	} & \multirow{3}{*}{ \begin{tabular}{l} \citet{Gaia2016} \\ \citet{Gaia2023} \end{tabular} \hfil }  \\
	& & & & \\
	& & & & \\
	% TODO: Calculate the actual lambda_iso with python code. 

	%\hline
	%$g_\text{APASS}$ & 4810 & \multirow{3}{*} { APASS } & \multirow{3}{*} {
	%\begin{tabular}{c}
	%	$\textrm{N / A}$\\
	%\end{tabular}
	%} & \multirow{3}{*} { \citet{Henden:2015aa} } \\
	%$r_\text{APASS}$ & 6170 & & & \\
	%$i_\text{APASS}$ &  7520 & & & \\

	%\hline
	%$g_\text{SDSS}$ & 4810 & \multirow{4}{*} { SDSS } & \multirow{4}{*} {
	%\begin{tabular}{c}
	%	$\textrm{N / A}$\\
	%\end{tabular}
	%} & \multirow{4}{*} { \citet{2017arXiv170709322A} } \\
	%$r_\text{SDSS}$ & 6170 & & & \\
	%$i_\text{SDSS}$ &  7520 & & & \\
	%$z_\text{SDSS}$ &  7520 & & & \\

	%\hline
	%$Z_\text{UKIDSS}$ & 8817  & \multirow{5}{*} { UKIDSS } & \multirow{5}{*} {
	%	\begin{tabular}{l}
	%	$\mathrm{pstar} > 0.99$\\$\mathrm{ j / h / k\_1 / x / ypperrbits} < 256$\\
	%	\end{tabular}
	%} & \multirow{5}{*} { \citet{2007MNRAS.379.1599L} } \\

	%$Y_\text{UKIDSS}$ & 10305  & & & \\
	%$J_\text{UKIDSS}$ & 12483 & & & \\
	%$H_\text{UKIDSS}$ & 16313 & & & \\
	%$K_{s, \text{UKIDSS}}$ & 22010 & & & \\

	\hline
	$J_\text{2MASS}$ & 12410 & \multirow{3}{*} { 2MASS } & \multirow{3}{*} { 
	\begin{tabular}{l}$\mathrm{ph\_qual} = \mathrm{"AAA"}$\\\end{tabular}
	} & \multirow{3}{*} { \citet{2mass}} \\
	$H_\text{2MASS}$ & 16513& & & \\
	$K_{s, \text{2MASS}}$ & 21643 & & & \\

	\hline
	\multirow{4}{*} {\begin{tabular}{l} $W_1$ \\ $W_2$ \\ $W_3$ \end{tabular}} &
	\multirow{4}{*} {\begin{tabular}{l} $33792$ \\ $46293$ \\ $123338$ \end{tabular}} &
	\multirow{4}{*} { WISE } & \multirow{4}{*} {
		\begin{tabular}{l}
		$\mathrm{ext\_flg} = 0$\\$\delta W_3 / W_3 < 0.05$ \\
		\end{tabular}
	} & \multirow{4}{*} {\citet{Wright:2010aa}} \\
	& & & & \\
	& & & & \\
	& & & & \\
	%$W_4$ & 222532 & & & \\
	\hline

	\end{tabularx}
	\label{tab:ch3-data-sources}
\end{table*}
We did not require that each of the stars were observed in all bands. 
However, we did require that each target was sampled by at least 5 bands, causing GJ 436 to be omitted from the final catalogue; owing to only having acceptable photometric data in the $G_{\rm BP}$ and $G_{\rm RP}$ bands. 
Thus our final sample contained 19 stars.
For reasons we will discuss in \autoref{sec:uv-photometry-discussion}, the NUV band was omitted from all SED fitting in this work, aside from in that section. 

% Address the WISE zero point here -- is this overkill?
We note that there is up to one per cent difference between different prescriptions in the WISE zero point; in particular the isophotal fluxes contained in Table 1 of \citet{Jarrett:2011aa} and the Vega spectrum contained in equation 2 of \citet{Wright:2010aa}.
However, this remains less than the known uncertainty in the WISE zero point of 1--4 per cent \citep{Jarrett:2011aa}. 
To probe the effect this uncertainty would have on our results, we performed fitting with the W3 zeropoint changed by 0.03 mag. 
Even in this pessimistic case, our final measured $R_{\star}$ remained within 1 per cent of the initial fitting; giving confidence that the uncertainty in the WISE zero point has minimal impact on our final results.

% section input-catalogue

\subsubsection{Extinction}
\label{sec:extinction}

Although most of our sample are within 100\ pc, some of the more distant stars are seen through significant interstellar extinction (with $V$-band extinctions of up to approximately 0.3 mag).
We therefore used the {\it Gaia} positions and parallaxes to place our targets in
the 10\ pc resolution dust maps of \cite{2022AA...664A.174V}.
We then integrated along the line of sight from the Sun to the star to calculate an extinction at 550\ nm [$A(55)$].
Since these maps were created using the extinction law of \cite{2019ApJ...886..108F} we used this law, combined with the mean Milky Way ratio of total-to-selective extinction ($R(55)=3.02$, corresponding to $R_V=3.1$), to  redden the models for each star.

\subsubsection{Photometric Fitting} % (fold)
\label{sec:photometric-fitting}
% MODEL GRID
We generated the grid for this work by folding the BT-Settl CIFIST stellar atmosphere models \citep{Allard:2012aa} through system responses corresponding to bands used in the fitting to produce synthetic photometry.
%Although the majority of discovered exoplanets orbit Solar-like stars, our inclusion of low-mass stars necessitated that we adopt models which incorporate thorough treatments of the molecular species which form in low-mass stellar atmospheres. 
The number of stars in the input catalogue is much smaller than that of our previous work, meaning that deriving uncertainties is computationally tractable for all targets. 
Hence, we performed a full 3D grid search in $T_{\rm eff}, \log(g)$ and $R_\star$ around the analytically determined radius for the entire catalogue. 
This produced a cube of $\chi^2$ in $T_{\rm eff} - \log(g) - R_\star$ space, from which the uncertainties for our measured parameters were drawn. 
The uncertainties were derived in accordance with our revised method in \citet{Morrell:2020aa}.
Using this method, we are able to estimate the $R_\star$ and $T_{\rm SED}$ bounds from even poorly sampled grids to within about 1 {per cent} of their true value.
% section photometric-fitting (end)

\subsubsection{Metallicity Corrections}
We demonstrated in \citet{Morrell:2019aa} that fitting non-Solar metallicity M-dwarf targets with only solar metallicity atmospheres induces an error in the measured parameters. 
In the same work we also proposed a strategy for obtaining corrected radii for M-dwarfs, which we will apply in this work. 
Equation (16) in \citet{Morrell:2019aa} provides the correction $\delta R_\star$ to be applied to the fitted radius as a function of luminosity $L_{\rm SED}$ and metallicity [Fe/H]. 
This mitigation exploits the fact that the SED fitting technique is able to measure luminosity accurately regardless of the disparity in metallicity between the target and model grid. 
This correction was applied to GJ 3470 and WASP-80, which are M-dwarfs for which we have reliable measurements of [Fe/H] from \citet{Demory:2013aa} and \citet{Triaud:2013aa} respectively. 
For both targets, we also corrected the presented $T_{\rm SED}$ value to remain consistent with the measured luminosity and revised radius. 
We will show in \autoref{sec:sources-of-systematic-error} that metallicity has a negligible effect on the measured parameters of Solar-type stars. 

\subsubsection{Stellar Parameters - Results} % (fold)
\label{sec:results}
%!TEX root = ../exoplanet-radii.tex

\begin{figure*}
\includegraphics[width=\textwidth]{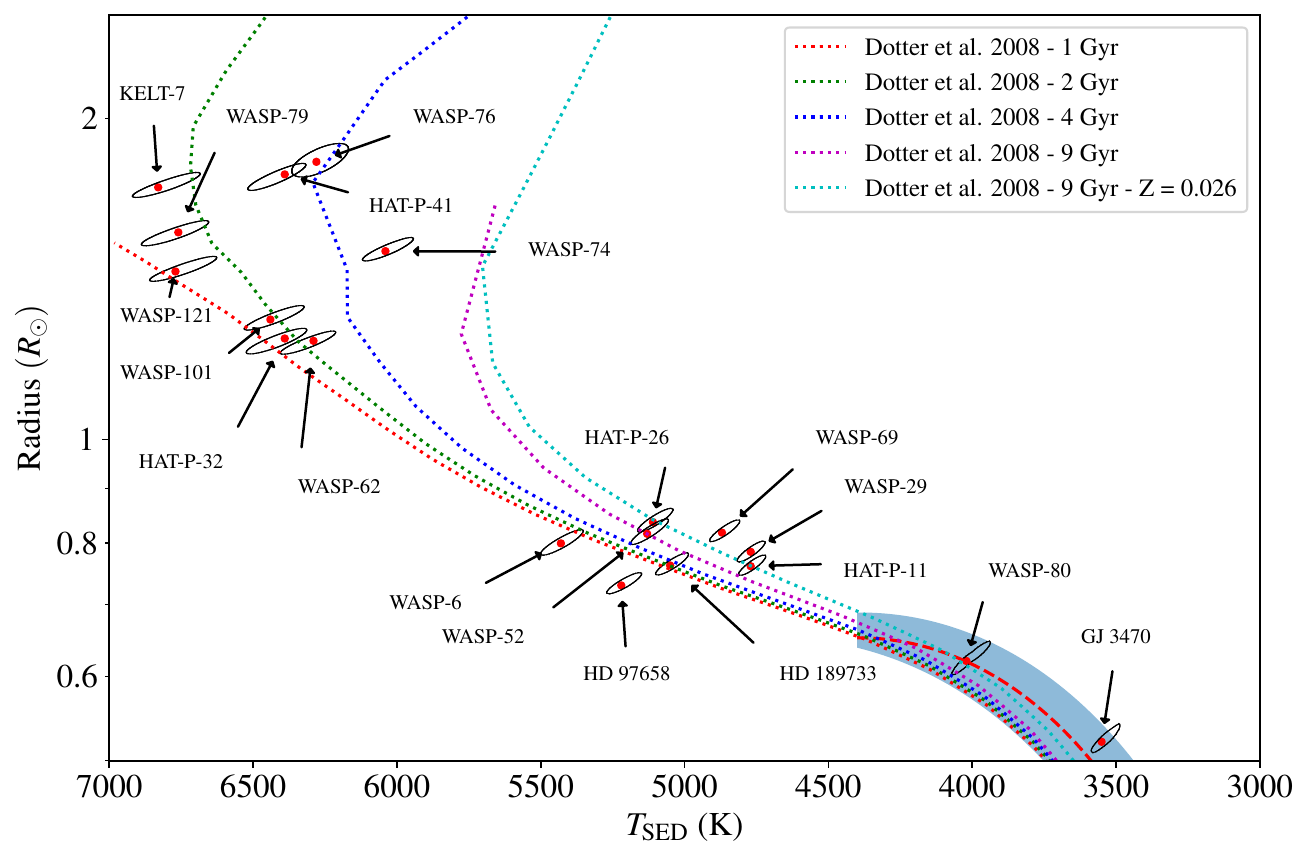}
\caption{
The host star properties in $T_{\rm SED} - R_\star$ space, along with their associated 68\% confidence contours. 
For comparison we show a series of isochrones from \citet{2008ApJS..178...89D} and the M-dwarf temperature radius relationship (red dashed line) derived in \citet{Morrell:2019aa} and its bounds (blue shaded region).
We emphasise that we do not expect all the stars to lie on these relationships, for example HD\,97658 which is known to be metal poor \citep{2011ApJ...730...10H} and so should lie below the Solar metallicity isochrones.
}

\label{fig:r-teff-plot-contour}
\end{figure*}

\begin{table}
	\centering
 	\caption{The stellar parameters $\log(g)$, $T_{\rm SED}$ and $R_*$ derived in this work from SED fitting. For the stars that require correction to the measured radius due to being non-Solar metallicity, both the $R_*$ and $T_{\rm SED}$ have been corrected in this table. 
    Whilst the fitting returns a value for $\log(g)$, in practice this is only weakly constrained by the SED.}
	\renewcommand{\arraystretch}{1.75}
	\begin{tabularx}{\columnwidth}{ l Y Y Y Y}
\hline
Name & $\log(g)$ & $T_{\rm SED}$ (K) & $R_*$ ($R_\odot$) \\\hline
GJ 3470 & $4.7^{+0.6}_{-0.3}$ & $3577^{+19}_{-51}$ & $0.513^{+0.014}_{-0.008}$\\ 
WASP-80 & $4.6^{+0.4}_{-0.2}$ & $4035^{+30}_{-62}$ & $0.616^{+0.020}_{-0.010}$\\ 
HAT-P-11 & $5.5^{+0.0}_{-0.7}$ & $4770^{+28}_{-40}$ & $0.761^{+0.013}_{-0.010}$\\ 
WASP-29 & $5.5^{+0.0}_{-0.7}$ & $4770^{+32}_{-36}$ & $0.785^{+0.012}_{-0.011}$\\ 
WASP-69 & $5.5^{+0.0}_{-0.7}$ & $4870^{+28}_{-49}$ & $0.818^{+0.017}_{-0.010}$\\ 
HD 189733 & $5.5^{+0.0}_{-0.9}$ & $5050^{+32}_{-48}$ & $0.762^{+0.015}_{-0.010}$\\ 
HAT-P-26 & $5.5^{+0.0}_{-0.9}$ & $5110^{+33}_{-51}$ & $0.837^{+0.018}_{-0.011}$\\ 
WASP-52 & $5.5^{+0.0}_{-0.9}$ & $5130^{+37}_{-56}$ & $0.816^{+0.019}_{-0.012}$\\ 
HD 97658 & $5.5^{+0.0}_{-0.9}$ & $5220^{+30}_{-52}$ & $0.730^{+0.015}_{-0.008}$\\ 
WASP-6 & $5.1^{+0.4}_{-1.1}$ & $5430^{+48}_{-58}$ & $0.799^{+0.016}_{-0.013}$\\ 
WASP-74 & $5.5^{+0.0}_{-0.9}$ & $6040^{+49}_{-70}$ & $1.502^{+0.031}_{-0.020}$\\ 
WASP-76 & $5.3^{+0.2}_{-0.8}$ & $6280^{+52}_{-80}$ & $1.821^{+0.050}_{-0.038}$\\ 
WASP-62 & $4.1^{+0.9}_{-0.1}$ & $6290^{+84}_{-48}$ & $1.237^{+0.015}_{-0.026}$\\ 
HAT-P-32 & $4.0^{+1.0}_{+0.0}$ & $6390^{+99}_{-46}$ & $1.244^{+0.016}_{-0.030}$\\ 
HAT-P-41 & $4.0^{+0.9}_{+0.0}$ & $6390^{+94}_{-43}$ & $1.772^{+0.024}_{-0.044}$\\ 
WASP-101 & $5.4^{+0.1}_{-0.9}$ & $6440^{+56}_{-87}$ & $1.296^{+0.028}_{-0.018}$\\ 
WASP-79 & $4.5^{+0.9}_{-0.1}$ & $6760^{+88}_{-71}$ & $1.564^{+0.025}_{-0.032}$\\ 
WASP-121 & $5.5^{+0.0}_{-0.9}$ & $6770^{+50}_{-109}$ & $1.437^{+0.033}_{-0.018}$\\ 
KELT-7 & $5.5^{+0.0}_{-0.9}$ & $6830^{+48}_{-111}$ & $1.723^{+0.039}_{-0.022}$\\ 
\hline
\end{tabularx}

	\label{tab:star-props}
\end{table}
The measurements of stellar properties resulting from the fitting process are presented with their uncertainties in \autoref{tab:star-props}. 
Our entire sample is shown in the $T_{\rm SED} - R_\star$ plane in Fig.\,\ref{fig:r-teff-plot-contour}, with the contours that denote their $68\%$ confidence region in this space. 
The mean uncertainties for the sample are 50\,K in temperature and 1.9 per cent in stellar radius.

\subsubsection{Testing the $G$-band magnitudes}
\label{sec:g_band_test}

\begin{figure}
	\includegraphics[width=\columnwidth]{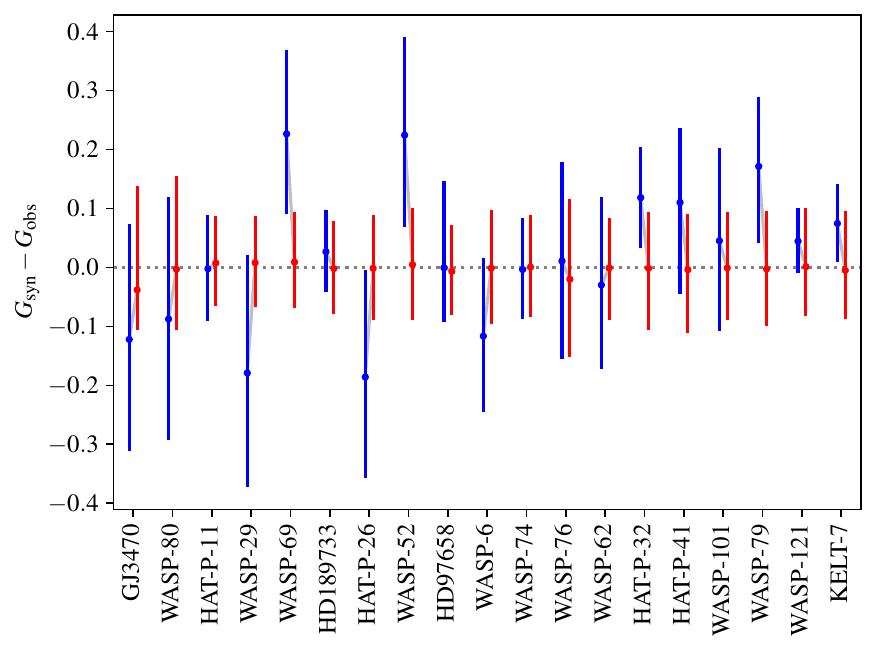}
	\caption{As Fig.\,\ref{fig:lit-lum-comp-names-litonly}, but with the added comparison between the synthetic $G$ band photometry generated from our parameters (red) shown alongside those generated from the literature values (blue). We have also estimated and plotted uncertainty bounds for each target. Note that these uncertainty bounds are likely over-estimates, as we did not have access to the correlation between the bounds in $T_{\rm eff}$ and $R_\star$. }
	\label{fig:lit-lum-comp-names}
\end{figure}

As an initial test of our new stellar parameters we return to the test we carried out in section \ref{sec:pancet_intro}.
In Fig.\,\ref{fig:lit-lum-comp-names} we have augmented Fig.\,\ref{fig:lit-lum-comp-names-litonly} with the equivalent synthetic $G$-band photometry resulting from an identical method, but using our revised $R_\star$ and $T_{\rm SED}$ presented in \autoref{tab:star-props}. 
In Fig.\,\ref{fig:lit-lum-comp-names}, we have also added uncertainty bounds for both sets of points. 
The lower uncertainty bound in $G$ was produced by adopting the lower uncertainty bound on $R_\star$ and $T_{\rm sp}$, and the upper uncertainty bound on distance $r_{\rm hi}$, and vice versa for the upper bound uncertainty. 
Hence these synthetic photometric uncertainties likely represent over-estimates of the true uncertainty since the temperature and radius are correlated.
This is borne out in that none of our measurements are inconsistent with the $G$ measurements---we would expect this to be the case for only $68\%$ of our measurements. 

Before addressing the comparison itself, we note that, at no point have we used the $Gaia$ $G$ band during the fitting process.
This is because its wide response, with an effective width $\Delta \lambda_{\rm eff}$ in excess of $4000$\,\AA, make it less effective for constraining visible colour than its narrower counterparts. 
However, especially for Solar-type stars where the majority of their radiant intensity is emitted at visible wavelengths, $G$ represents a robust proxy for visible stellar luminosity. 
Hence, agreement between the observed $G$ magnitude and the synthetic photometry resulting from our revised parameters is not guaranteed by the fitting, and should instead be interpreted as a strong indication that our luminosity measurements are accurate.  

% Uncertainty bounds
There are some important points to be drawn from this comparison. 
First, we note that the comparison between the observed $G$-band data and the synthetic photometry resulting from our parameters exhibit a factor of 10 reduction in RMS residual in $G$. 
Furthermore, compared to the literature measurements, our dataset exhibit nearly a 40 {per cent} improvement in the mean uncertainty in $G$. 
Despite this, it is interesting to note that our values of $T_{\rm SED}$ are in good agreement with the measured literature values of $T_{\rm sp}$, with 17 of the 19 $T_{\rm SED}$ measurements being within $2 \sigma$ of literature $T_{\rm sp}$ values. 

\subsubsection{Validating the Stellar Temperatures}
\label{sec:test_teff}
\begin{figure}
	\includegraphics[width=\columnwidth]{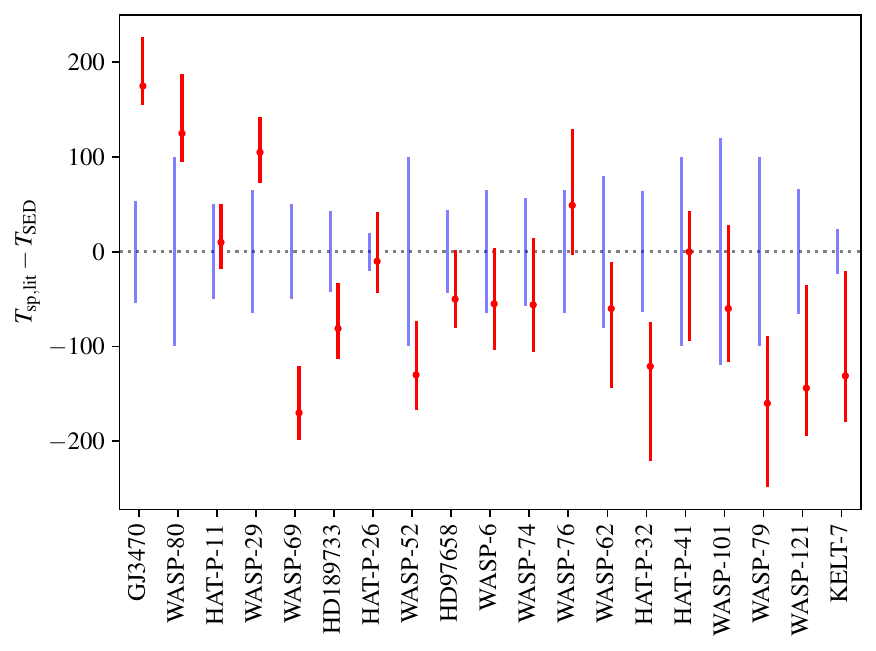}
	\caption{A comparison between the $T_{\rm SED}$ measured in this work and the $T_{\rm sp}$ presented in the exoplanet literature for all of the targets in our sample. 
    The red points are the temperature differences between our $T_{\rm SED}$ measurements and the literature $T_{\rm sp}$ measurements.
    The associated red error bars are our uncertainties.
    The blue error bars (centred on zero) represent the literature uncertainties, and thus for consistent measurements the red and blue error bars should overlap.}
	\label{fig:lit-teff-comp-names}
\end{figure}
\begin{figure*}
	\includegraphics[width=\textwidth]{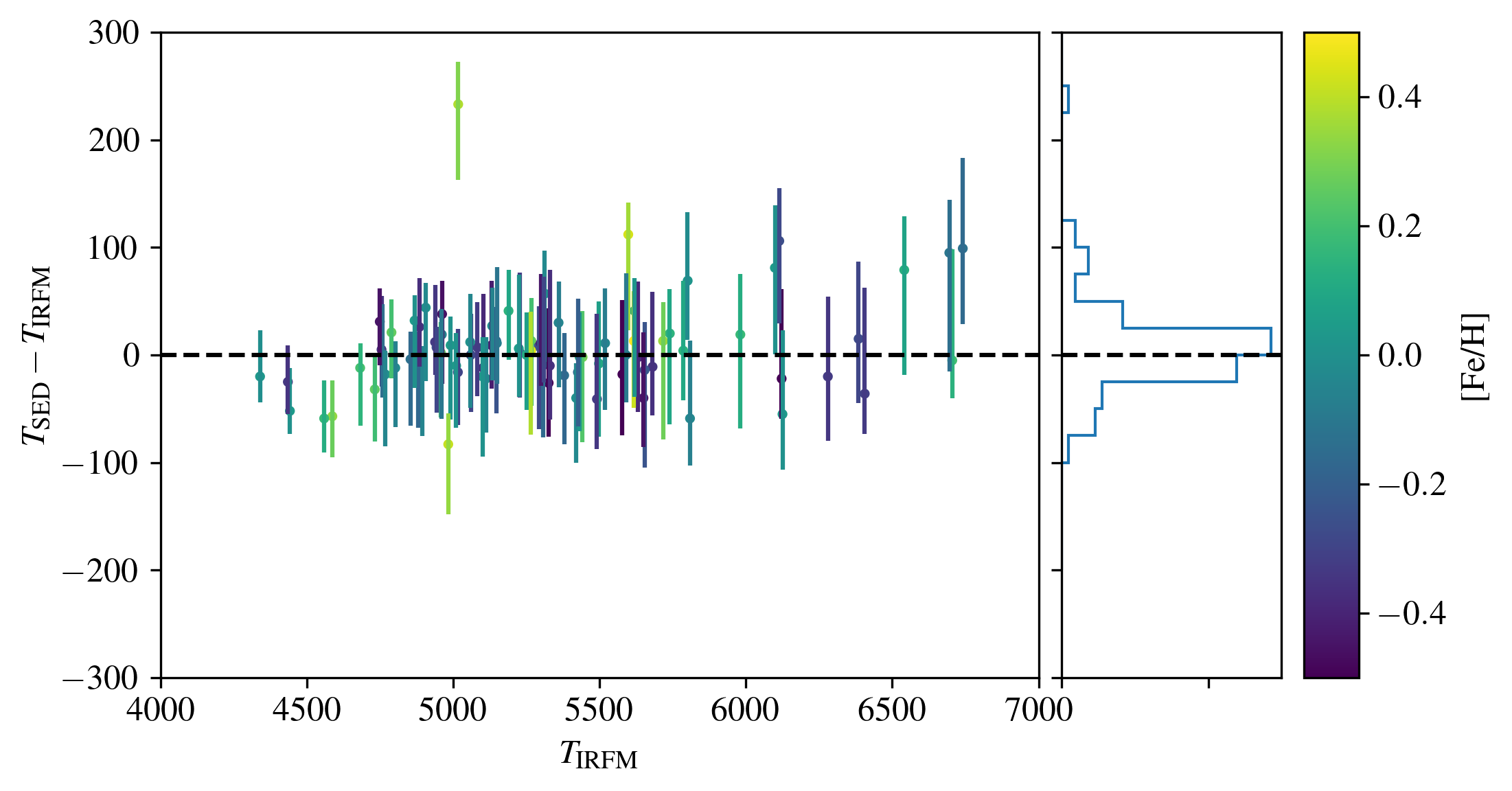}
	\caption{A comparison between the $T_{\rm SED}$ measured in this work and the infrared flux method temperature measurements of \citet{2010A&A...512A..54C}. The colour of each data point denotes the metallicity of the source, with the errorbars denoting the 68\% confidence bounds of our measured $T_{\rm SED}$. A histogram of the difference between $T_{\rm SED}$ and $T_{\rm IRFM}$ is shown in the auxilliary subplot.
    }
	\label{fig:casagrande2010-comparison}
\end{figure*}
\begin{table*}
    \centering
    \begin{tabular}{lccccc}
    \hline
            &           & Mean                      & RMS about & Mean $T_{\rm SED}$    \\
            & Number    & $T_{\rm SED}-T_{\rm eff}$ & mean      & Uncertainty \\
Comparator  & of points & (Kelvin)   & (Kelvin)  & (Kelvin) & $\chi^2_{\nu}$ \\
\hline
Casagrande at al (2010) & 87  &   3 $\pm$ 8 &  44 &  52 &  0.25 \\
Solar twins and analogues & 40 & 11 $\pm$ 9 & 43 & 57 & 0.61 \\
Brewer et al (2016) & 123 & 48 $\pm$ 5 & 43 & 54 & 0.54 \\
Freckelton et al (2025) & 546 & 23 $\pm$ 5 & 87 & 56 & 0.52 \\
\hline
    \end{tabular}
    \caption{The results from a weighted mean analysis of the residuals between the SED temperatures and various literature sources.  The \citet{2016ApJS..225...32B} and \citet{2025MNRAS.540.1786F} comparisons are for temperatures cooler than 6\,000\,K. }
    \label{tab:comparison_fits}
\end{table*}

In Fig.\,\ref{fig:lit-teff-comp-names} we show the differences between the literature temperatures and our $T_{\rm SED}$ values for the PanCET sample.
This shows an overall agreement, with our uncertainties being significantly smaller.
However, there is no way to test our uncertainties with this sample, and although there appears to be no significant systematic shift of the temperature scale, the sample is small.
Therefore, as a further test of our temperatures we compared them with four literature sources.
In all cases we only used stars whose SEDs fits had five or more data points and $\chi^2$ values less than ten, and had [Fe/H] within 0.5 dex of solar according to the values given in the literature source used for the comparison.
Finally in all cases except the sample of Solar twins and analogues we removed stars whose SIMBAD \citep{2000A&AS..143....9W} object types indicated possible variability or binarity.\footnote{We removed objects whose $Gaia$ DR3 identifiers were linked by SIMBAD to primary or other object types **,  Ae*,  BYDraV*, EclBin, Eruptive*, RGB*, RSCVnV*,  SB*, Variable*, YSO or  delSctV*; or their shortform equivalents.
We also removed $Gaia$ DR3 1186325496788069632 which is not in SIMBAD but appears to be the binary star HD131473.}

We calculated uncertainties for the temperature differences by adding the error bars from  both samples in quadrature, using the mean of our upper and lower uncertainties for the SED temperatures. 
The results of these comparisons are given in Table \ref{tab:comparison_fits}, where we report the results of fitting a constant to the temperature differences (mathematically equivalent to calculating the weighted mean), including a reduced $\chi^2$ assuming a one-free-parameter fit.

We first compared our technique to the Infrared Flux Method (IRFM) temperatures of \cite{2010A&A...512A..54C}.
We calculated SED temperatures for their  sample, but only retained those for which our 68 per cent confidence region for $T_{\rm SED}$ lay entirely below 7\,000\,K (the hottest atmosphere in our grid), and above 3\,000\,K.
The mean temperature difference is zero to within our uncertainties and Fig.\,\ref{fig:casagrande2010-comparison} shows this is probably true throughout the temperature range, though below 4\,750\,K the mean difference is around $-$40\,K.
There is no obvious reason for this; for example those stars are not significantly different in [Fe/H].
In addition our data are sparse above 6\,500\,K, and so a safe statement to make is that the mean difference between our temperature scale and that of \cite{2010A&A...512A..54C} is less than 5\,K for 4\,750\,K $<$ $T_{\rm eff}$ $<$ 6\,500\,K.

Although this comparison suggests $T_{\rm SED}$ provides a robust and accurate temperature scale, it is worth noting that by comparing it with an IRFM scale we are comparing techniques whose input data are qualitatively similar; both use broadband photometry.
However, the methodology used to analyse those data is different (see Appendix \ref{sec:history}), and with the exception of 2MASS the data are different, as are the model atmospheres.

In contrast to the IRFM, spectroscopically determined temperatures allow us to compare $T_{\rm SED}$ against an entirely independent technique.
The highest precision temperatures available are those determined for Solar twins and analogues by differential techniques with respect to the Sun.
\cite{Spina:2018aa} and 
\cite{2023MNRAS.522.3217M} derive such temperatures with uncertainties of a few Kelvin.
We examined the sample of these stars which overlap with the $T_{\rm SED}$ values we presented in \cite{2025MNRAS.540.1786F}, which are calculated using identical software to the PanCET sample presented here.
Again the mean difference in temperature is consistent with zero.

\begin{figure*}
    \centering
    \includegraphics[width=\linewidth]{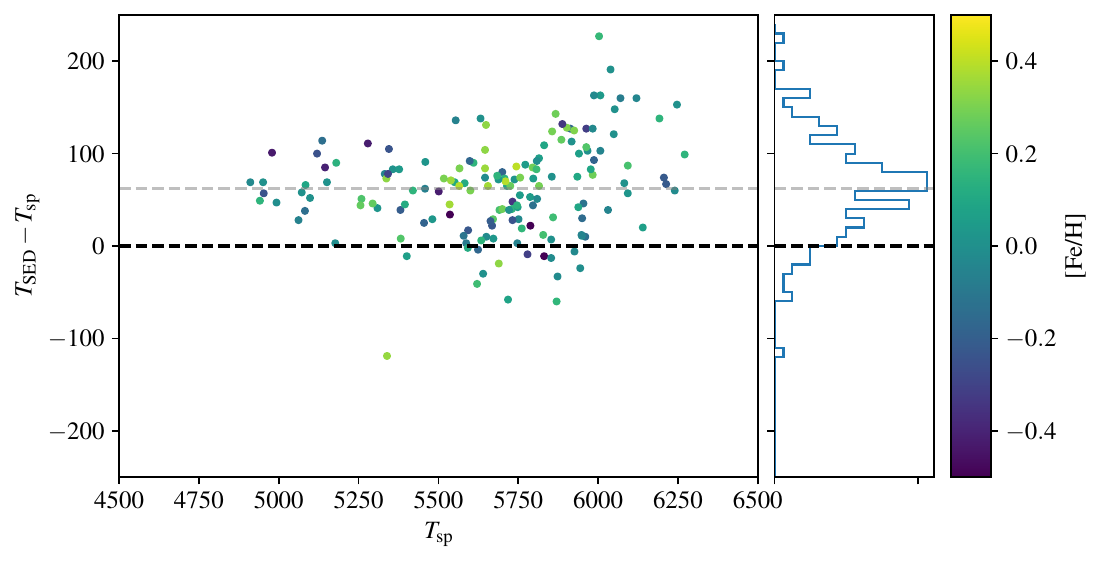}
    \caption{A comparison between the $T_{\rm SED}$ measured in this work and the spectroscopic temperatures presented in \citet{2016ApJS..225...32B}. Note that 4 matched targets fall outside the $y$-axis extent of this plot. As in \autoref{fig:casagrande2010-comparison}, the colour of each data point denotes the metallicity of the target. The mean temperature difference is overlaid as a grey dashed line. A histogram of the differences between $T_{\rm SED}$ and $T_{\rm sp}$ is shown in the auxiliary subplot.  }
    \label{fig:brewer2016_comparison}
\end{figure*}

By construction the sample of Solar twins and analogues only covers a range of $\pm 200$\,K, so we also compared our temperatures with the spectroscopic determinations of \cite{2016ApJS..225...32B}. 
Once again we used the $T_{\rm SED}$ values presented in \cite{2025MNRAS.540.1786F}.
\autoref{fig:brewer2016_comparison} shows that in addition to a mean offset of about 50\,K, above 6\,000\,K there is a significant increase in $T_{\rm SED}$ compared with these spectroscopic temperatures.
\cite{2025MNRAS.540.1786F} performed a similar comparison, comparing $T_{\rm SED}$ with the spectroscopic temperatures also presented in that paper.
The result has a similar rise in the residuals beyond 6\,000\,K, though a smaller mean offset (see also Table \ref{tab:comparison_fits}).

The best comparison for exploring our uncertainties is the sample of Solar twins and analogues as their uncertainties are small compared with those from $T_{\rm SED}$, and hence our uncertainties completely dominate the combined error bar.
The $\chi^2_{\nu}$ of the fit to the residuals is 0.61, suggesting our uncertainties might be over-estimated by a factor of about 1.2.
The other comparisons follow the same pattern, with low values of $\chi^2_{\nu}$, though of course in principle this could be due to other authors over-estimating their uncertainties.
This suggestion that our uncertainties are too high is also reflected in the low $\chi^2$ values of our SED fits \citep[see the online table of][]{2025MNRAS.540.1786F}, which imply that the photometric uncertainties we used are too large.
This could be due to the one percent floor we impose on them to allow for uncertainties in the zero point.
The implication is that our true parameter uncertainties may be smaller than quoted here, if the systematic errors in the photometric zero points are smaller than currently believed.

In summary it appears our estimates of the random uncertainties for our temperatures are correct, or perhaps even over-estimates. 
However, the offsets with respect some to temperature scales suggest there may be a systematic of several tens of Kelvin, which increases beyond 6\,000\,K.
Such offsets are not unusual \citep[see, for example, the discussion in][]{2016ApJS..225...32B}, but their causes are unclear.

\subsubsection{Validating the Stellar Radii}
\label{sec:test_rad}
\begin{figure}
    \centering
    \includegraphics[width=\columnwidth]{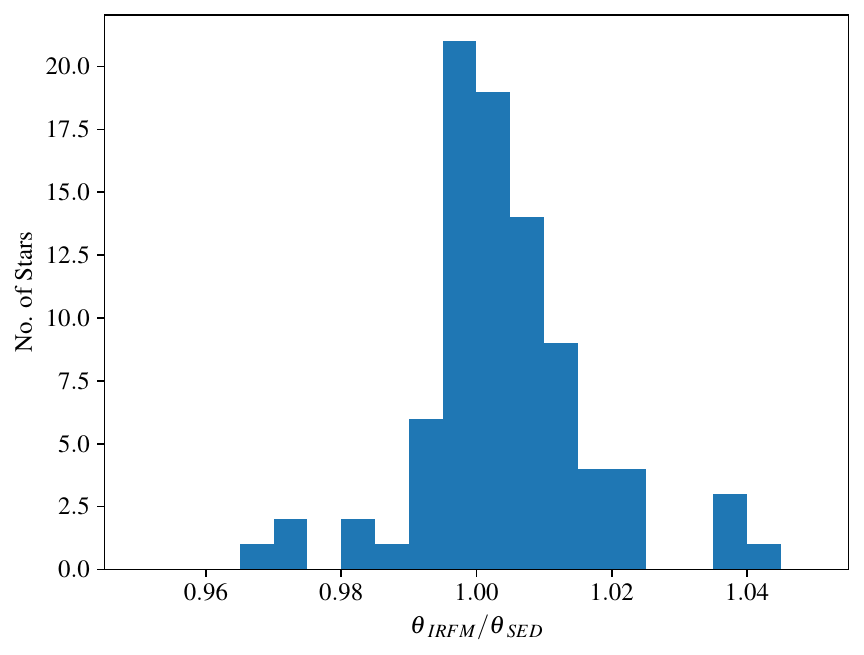}
    \caption{A histogram of the ratio of the angular radii of stars measured using the Infrared Flux Method by \citet{2010A&A...512A..54C} to those measured in this work.  The sample is limited to stars with -0.5 < [Fe/H] < +0.5. } 
    \label{fig:casagrande2010-angdiam-comparison}
\end{figure}

We can test our stellar radii against those of \cite{2010A&A...512A..54C} in a similar way to that used for the temperatures, albeit with the same reservations about the similarity of technique.
The comparison is best carried out using angular diameters, as that is how the data are presented in \cite{2010A&A...512A..54C}, and any conversion to stellar radius would use the same $Gaia$ distances for both datasets.
Fig.~\ref{fig:casagrande2010-angdiam-comparison} shows the comparison for the same sample of stars used in section \ref{sec:test_teff}.
The mean ratio is 1.0039, with an RMS of 0.013, which gives an uncertainty for this mean of 0.0014.
Hence our radii agree to within about 0.5 per cent.
%There is some evidence for systematic deviations with temperature from this mean, in particular a systematic increase of about 2 per cent for stars cooler than 4\,750\,K.
%This exactly mirrors the 50\,K (1 per cent) systematic shift for temperatures. 
%The anti-correlation reflects the fact that cooler atmospheres require larger radii to match the observed luminosity, and the relative sizes of the changes that for fixed luminosity the radius varies as the temperature squared.
We know that the \cite{2010A&A...512A..54C} radii link (indirectly) to the interferometric measurements to within about 2 per cent, giving confidence that our measurements are also close to the most directly measured stellar radii available.
The median uncertainty we derive for the angular diameters is 2.4 per cent.
However, the reduced $\chi^2$ of the data in Fig.~\ref{fig:casagrande2010-angdiam-comparison} with respect to 1.0 is 0.29, implying our error bars for angular diameter and hence radius are, like those for temperature, significantly overestimated.

\subsection{Stellar Parameter Discussion}
\label{sec:stellar-discuss}
\begin{figure*}
	\includegraphics[width=\textwidth]{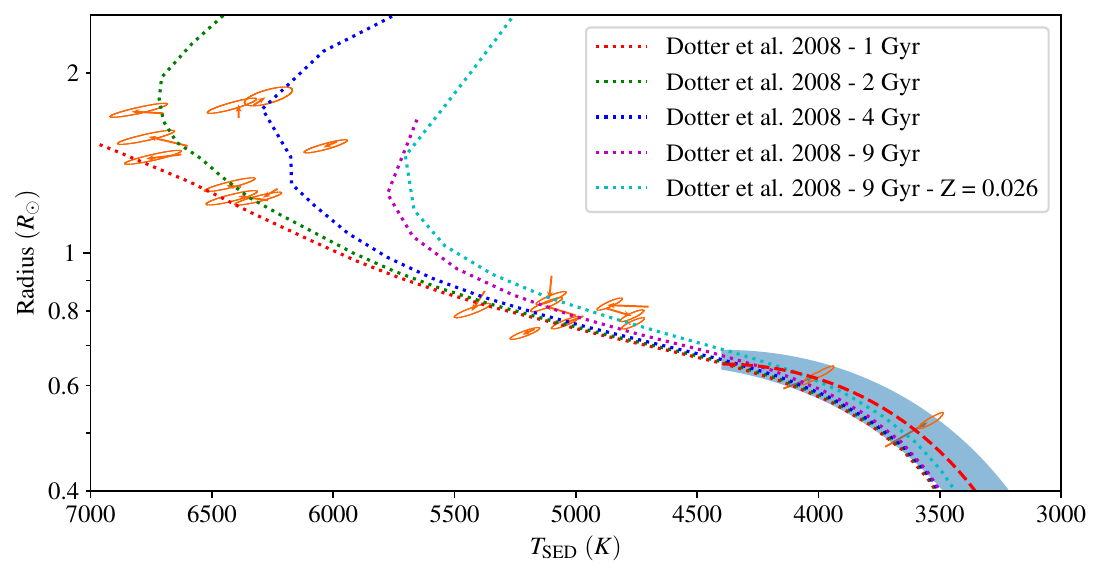}
	\caption{Vectors mapping the literature value onto the revised values in the $T_{\rm SED} - R_\star$ plane. The arrows show the vector that maps the literature value from the PanCET primary target list onto the value measured in this work, using the SED fitting method. The green denotes HD 189733, for which the literature radius was determined using interferometry. Those remaining, shown in orange, are derived from the light curves and stellar models. }
	\label{fig:vector-comp}
\end{figure*}
We first performed a comparison between our measurements and those from the literature in the $T_{\rm SED} / T_{\rm sp} - R_\star$ plane in Fig.\,\ref{fig:vector-comp}. 
This figure demonstrates that there is no clear systematic difference between the radii and temperatures presented in the literature and those measured from SED fitting with respect to the main sequence. 

\begin{figure}
\includegraphics[width=\columnwidth]{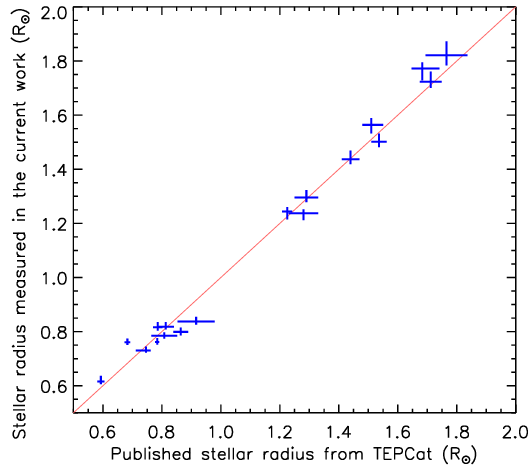} \\
\caption{\label{fig:R1} Comparison between the host star radii measured
in the literature and our own results. The red line is parity.}
\end{figure}
Fig.\,\ref{fig:R1} shows that the agreement between our calculated radii for the host stars and those from the literature is good. 
As in section \ref{sec:pancet_intro} the literature values were taken from TEPCat, though they mostly correspond to values given in the publications cited in Table\,\ref{tab:input}. 

\subsection{Exoplanet Parameters}
\label{sec:planet_params}

With the stellar radii determined, we can now move to measure the mass and radius of the exoplanets. Our approach is similar to that from previous works such as \citet{2018ApJ...862...53S} and \citet{2024MNRAS.530.2565D}, but is described in full for clarity and convenience

\subsubsection{The method}

The determination of the physical properties of extrasolar planets and their host stars has typically been complicated by having insufficient information to do so algebraically. In terms of an analogy with eclipsing binary systems \citep[e.g.][]{Hilditch01book}, the velocity semiamplitude of the secondary component (i.e.\ the planet) is not measured and so the full scale of the system and the properties of its components cannot be calculated directly \citep[see][]{SeagerMallen03apj,Torres++08apj,Me09mn}. This issue is usually circumvented by forcing the properties of the host star to match the predictions of theoretical evolutionary models, thus imprinting a model-dependence on the outcome.

With the directly-measured $T_{\rm eff}$ and radius ($R_\star$) values for the host stars determined in this work, we are able to avoid this problem entirely. The bulk physical properties of the system can be found from these two measurements plus parameters measured directly from the light curve of the system and radial velocity curve of the host star. The required parameters are the orbital period ($P$), eccentricity ($e$) and inclination ($i$) of the system, the density ($\rho_\star$) and velocity semiamplitude ($K_\star$) of the star and the ratio of the radii ($k = R_{\rm p} / R_\star$).

The first step is to extract a value for $\rho_\star$ from the light curve. The parameters of a light curve that can be obtained are effectively $P$, $i$, and the fractional radii of the components ($r_\star = R_\star/a$ and $r_{\rm p} = R_{\rm p}/a$ where $a$ is the semimajor axis of the relative orbit) \citep{Russell12apj}. Complications such as limb darkening and eccentricity must be accounted for, and both are the subject of extensive literature. Using Kepler's third law and the definition of density it can be shown that:
\begin{equation}
\rho_\star = \frac{3\pi}{GP^2}\frac{M_\star}{M_\star+M_{\rm p}}\left(\frac{a}{R_\star}\right)^3 ~,
\end{equation}
where $G$ is the gravitational constant, $M_\star$ is the stellar mass and $M_{\rm p}$ is the planetary mass. If we make the safe assumption that $M_{\rm p} \ll M_\star$ and convert to fractional radius we find:
\begin{equation} \label{eq:rho1}
\rho_\star \approx \frac{3\pi}{GP^2r_\star^{\,3}} ~,
\end{equation}
and thus $\rho_\star$ is determinable from the light curve alone in transiting planetary systems.

Armed with $R_\star$ and $\rho_\star$ in solar units the stellar mass can be found from the definition of density:
\begin{equation} \label{eq:M1}
\left(\frac{M_\star}{\rm M_\odot}\right) = \left(\frac{\rho_\star}{\rm \rho_\odot}\right) \left(\frac{R_\star}{\rm R_\odot}\right)^3 ~.
\end{equation}
The surface gravity of the star follows from the standard formula based on the Sun:
\begin{equation}
\log g_\star ({\rm c.g.s.}) = 4.438 + \log\left(\frac{M_\star}{\rm M_\odot}\right) - 2\log\left(\frac{R_\star}{\rm R_\odot}\right) ~.
\end{equation}
The mass function can then be found from:
\begin{equation}
f(M) = \frac{P (1-e^2)^{3/2} K_\star^{\,3}}{2 \pi G} ~.
\end{equation}
The mass function can also be expressed in a way involving $M_{\rm p}$:
\begin{equation} \label{eq:fm}
f(M) = \frac{M_{\rm p}^{\,3} \sin^3i}{(M_\star+M_{\rm p})^2} ~,
\end{equation}
which can be solved in two ways to determine $M_{\rm p}$ from $M_\star$ and $f(M)$. The first way is to solve Eq.\,\ref{eq:fm} iteratively, which in practise converges quickly and reliably for planetary systems because $M_{\rm p} \ll M_\star$. The second way is to \emph{assume} $M_{\rm p} \ll M_\star$ and thus rewrite Eq.\,\ref{eq:fm} as
\begin{equation} \label{eq:fm2}
M_{\rm p} \sin i \approx M_\star^{~2/3} f(M)^{1/3} ~.
\end{equation}
We chose the first option in the current case, to avoid the approximation and because implementation is straightforward. Once both masses are known, the semimajor axis of the relative orbit follows from Kepler's third law via:
\begin{equation}
a = \left(\frac{G(M_\star+M_{\rm p})P}{4\pi^2}\right)^{1/3} ~.
\end{equation}

Now it is possible to determine the properties of the planet, beginning with the planetary radius:
\begin{equation}
R_{\rm p} = k R_\star \,.
\end{equation}
Its surface gravity can be determined purely from observed quantities \citep{Me++07mn} using:
\begin{equation}
g_{\rm p} = \frac{GM_{\rm p}}{R_{\rm p}^{\,2}} = \frac{2\pi}{P}\frac{\sqrt{1-e^2}}{r_{\rm p}^{\,2}\sin i}K_\star ~.
\end{equation}
The density can be found from
\begin{equation}
\left(\frac{\rho_{\rm p}}{\rm \rho_{\rm Jup}}\right) = \left(\frac{M_{\rm p}}{\rm M_{\rm Jup}}\right) \left(\frac{R_{\rm p}}{\rm R_{\rm Jup}}\right)^{-3} ~,
\end{equation}
and the equilibrium temperature from
\begin{equation}
T_{\rm eq} = T_{\rm eff} \sqrt{\frac{R_\star}{2a}}
\end{equation}
where for convenience we have neglected scaling factors due to albedo and heat redistribution \citep[e.g.][]{Guillot+96apjl}.

\subsubsection{Exoplanet Parameters - Results}
\begin{table*} 
\caption{Summary of the input parameters for determining the exoplanet properties and stellar masses.
A quantity in brackets represents the uncertainty in the last digit of the preceding value.
Some of the eccentricity values are upper limits.}
\renewcommand{\arraystretch}{1.5}
\begin{tabular}{lccccccl} \hline
System    &      $P$ (d)   & $K_\star$ (m\,s$^{-1}$)    &            $e$              & $\rho_\star$ ($\rho_\odot$) &              $k$               &   $i$ ($^\circ$)       & References      \\
\hline
GJ 3470   & 3.3366527  (3) & \er{  8.03}{ 0.38}{ 0.37}  & \er{0.125 }{0.043 }{0.042 } & \er{3.39  }{0.30  }{0.32  } & \er{0.0777 }{0.0026 }{0.0026 } & \er{88.14}{0.82}{0.64} & 1, 2, 3, 4      \\
WASP-80   & 3.0678569 (16) & \er{110.9 }{ 3.0 }{ 3.3 }  & \er{0.0   }{0.07  }{0.0   } & \er{2.862 }{0.050 }{0.050 } & \er{0.17058}{0.00057}{0.00057} & \er{89.92}{0.07}{0.12} & 5, 6            \\
HAT-P-11  & 4.8878024  (1) & \er{ 10.2 }{ 1.1 }{ 1.2 }  & \er{0.2646}{0.0007}{0.0005} & \er{1.915 }{0.081 }{0.081 } & \er{0.05860}{0.00037}{0.00037} & \er{89.36}{0.36}{0.36} & 7, 8, 9, 10, 11 \\
WASP-29   & 3.9227122  (2) & \er{ 35.6 }{ 2.7 }{ 2.7 }  & \er{0.03  }{0.05  }{0.03  } & \er{1.68  }{0.07  }{0.15  } & \er{0.09668}{0.00080}{0.00080} & \er{89.32}{0.68}{0.68} & 12, 13          \\
WASP-69   & 3.8681358  (3) & \er{ 38.1 }{ 2.4 }{ 2.4 }  & \er{0.0   }{0.10  }{0.0   } & \er{1.54  }{0.13  }{0.13  } & \er{0.1336 }{0.0016 }{0.0016 } & \er{86.71}{0.20}{0.20} & 14              \\
HD 189733 & 2.2185752  (1) & \er{200.56}{ 0.88}{ 0.88}  & \er{0.0041}{0.0025}{0.0020} & \er{1.625 }{0.094 }{0.094 } & \er{0.15565}{0.00024}{0.00021} & \er{85.71}{0.01}{0.01} & 15, 16, 17, 18  \\
WASP-52   & 1.7497815  (4) & \er{ 84.3 }{ 3.0 }{ 3.0 }  & 0.0 fixed                   & \er{1.653 }{0.020 }{0.020 } & \er{0.16378}{0.00050}{0.00050} & \er{85.05}{0.06}{0.06} & 19, 20          \\
HAT-P-26  & 4.23450236 (3) & \er{  8.5 }{ 1.0 }{ 1.0 }  & \er{0.124 }{0.060 }{0.060 } & \er{1.04  }{0.22  }{0.22  } & \er{0.0732 }{0.0011 }{0.0011 } & \er{88.09}{0.55}{0.55} & 21, 22          \\
HD 97658  & 9.4893077  (2) & \er{  2.73}{ 0.26}{ 0.27}  & \er{0.078 }{0.057 }{0.053 } & \er{1.89  }{0.23  }{0.20  } & \er{0.0293 }{0.001  }{0.001  } & \er{89.6 }{0.1 }{0.1 } & 23, 24, 25      \\
WASP-6    & 3.3610021  (3) & \er{ 74.3 }{ 1.7 }{ 1.4 }  & \er{0.054 }{0.018 }{0.014 } & \er{1.296 }{0.053 }{0.053 } & \er{0.1463 }{0.0012 }{0.0012 } & \er{88.38}{0.31}{0.31} & 26, 27          \\
WASP-74   & 2.1377445  (7) & \er{114.1 }{ 1.4 }{ 1.4 }  & \er{0.0   }{0.07  }{0.0   } & \er{0.329 }{0.012 }{0.012 } & \er{0.09034}{0.00063}{0.00063} & \er{79.86}{0.21}{0.21} & 28, 29, 30      \\
HAT-P-41  & 2.6940486  (9) & \er{ 92.5 }{11.6 }{11.6 }  & 0.0 fixed                   & \er{0.297 }{0.015 }{0.026 } & \er{0.1028 }{0.0016 }{0.0016 } & \er{87.7 }{0.1 }{0.1 } & 31              \\
HAT-P-32  & 2.1500082  (1) & \er{122.8 }{23.2 }{23.2 }  & 0.0 fixed                   & \er{0.6435}{0.0032}{0.0032} & \er{0.1515 }{0.0004 }{0.0004 } & \er{88.98}{0.21}{0.21} & 32, 33          \\
WASP-76   & 1.8098820  (6) & \er{119.3 }{ 1.8 }{ 1.8 }  & \er{0.0   }{0.05  }{0.0   } & \er{0.2722}{0.0057}{0.0057} & \er{0.10738}{0.00018}{0.00018} & \er{87.29}{0.50}{0.50} & 34, 13          \\
WASP-62   & 4.4119386  (6) & \er{ 61   }{ 3   }{ 3   }  & \er{0.0   }{0.21  }{0.0   } & \er{0.59  }{0.06  }{0.06  } & \er{0.1111 }{0.0001 }{0.0001 } & \er{88.5 }{0.4 }{0.7 } & 35, 36, 37      \\
WASP-101  & 3.585722   (4) & \er{ 54   }{ 4   }{ 4   }  & \er{0.0   }{0.03  }{0.0   } & \er{0.626 }{0.043 }{0.043 } & \er{0.10860}{0.00019}{0.00019} & \er{84.73}{0.05}{0.05} & 38, 39          \\
WASP-121  & 1.27492471 (8) & \er{177.0 }{ 8.5 }{ 8.1 }  & \er{0.0   }{0.07  }{0.0   } & \er{0.437 }{0.008 }{0.009 } & \er{0.12534}{0.00043}{0.00060} & \er{88.49}{0.16}{0.16} & 40, 41          \\
WASP-79   & 3.6623926  (5) & \er{ 88.5 }{ 7.7 }{ 7.7 }  & 0.0 fixed                   & \er{0.3658}{0.0085}{0.0085} & \er{0.10680}{0.00021}{0.00021} & \er{85.60}{0.13}{0.13} & 42, 13          \\
KELT-7    & 2.7347655  (3) & \er{140   }{17   }{17   }  & 0.0 fixed                   & \er{0.301 }{0.025 }{0.025 } & \er{0.08957}{0.00012}{0.00012} & \er{83.51}{0.09}{0.09} & 43, 44          \\
\hline
\label{tab:input}
\end{tabular}
\begin{flushleft}
{\bf References:}
 1 \citet{Bonfils+12aa},
 2 \citet{Biddle+14mn},
 3 \citet{Chen+17aa},
 4 \citet{Stefansson+22apj},
 5 \citet{Triaud+13aa},
 6 \citet{Mancini+14aa},
 7 \citet{Bakos+10apj},
 8 \citet{Me11mn},
 9 \citet{Knutson+14apj},
10 \citet{Huber+17aa},
11 \citet{Basilicata+24aa},
12 \citet{Hellier+10aa},
13 This work,
14 \citet{Anderson+14mn},
15 \citet{Bouchy+05aa},
16 \citet{Boisse+09aa},
17 \citet{Triaud+09aa},
18 \citet{Krenn+23aa},
19 \citet{Hebrard+13aa},
20 \citet{Mancini+17mn},
21 \citet{Hartman+11apj2},
22 \citet{Mancini+22aa},
23 \citet{Dragomir+13apj},
24 \citet{Vangrootel+14apj},
25 \citet{Guo+20aj},
26 \citet{Gillon:2009aa},
27 \citet{Tregloan+15mn},
28 \citet{Hellier+15aj},
29 \citet{Mancini+19mn},
30 \citet{Spyratos+23mn},
31 \citet{Hartman:2012aa},
32 \citet{Hartman+11apj},
33 \citet{Tregloan+18mn},
34 \citet{West+16aa},
35 \citet{Hellier+12mn},
36 \citet{Brown+17mn},
37 \citet{PatelEspinoza22aj},
38 \citet{Hellier+14mn},
39 \citet{Wakeford+17apj},
40 \citet{Delrez+16mn},
41 \citet{Bourrier+20aa},
42 \citet{Smalley+12aa},
43 \citet{Bieryla+15aj},
44 \citet{Tabernero+22mn}.
\end{flushleft}
\end{table*}

\begin{table*} 
\caption{Physical properties of the transiting planetary systems calculated in this work.}
\renewcommand{\arraystretch}{1.5}
\begin{tabular}{lcccccccc} \hline
System    &            $a$ (au)         &     $M_\star$ (\Msun)    &      $\log g_\star$      &    $M_{\rm p}$ (\Mjup)   &   $R_{\rm p}$ (\Rjup) & $g_{\rm p}$ (m\,s$^{-2}$) &  $\rho_{\rm p}$ (\pjup)  & $T_{\rm eq}$ (K)  \\
\hline
GJ 3470   & \er{0.03367}{0.00135}{0.00113} & \er{0.457}{0.057}{0.044} & \er{4.678}{0.040}{0.041} & \er{0.035}{0.003}{0.003} & \er{0.388}{0.017}{0.013} & \er{ 5.73}{0.59}{0.58} & \er{0.596}{0.074}{0.076} & \er{ 673}{10}{13} \\
WASP-80   & \er{0.03612}{0.00117}{0.00064} & \er{0.668}{0.067}{0.035} & \er{4.684}{0.016}{0.010} & \er{0.606}{0.041}{0.028} & \er{1.022}{0.033}{0.017} & \er{14.39}{0.41}{0.48} & \er{0.568}{0.018}{0.027} & \er{ 803}{ 6}{12} \\
HAT-P-11  & \er{0.05325}{0.00116}{0.00099} & \er{0.843}{0.056}{0.046} & \er{4.601}{0.019}{0.019} & \er{0.073}{0.008}{0.009} & \er{0.434}{0.008}{0.006} & \er{ 9.66}{1.06}{1.15} & \er{0.898}{0.098}{0.110} & \er{ 869}{ 7}{ 9} \\
WASP-29   & \er{0.04541}{0.00090}{0.00150} & \er{0.812}{0.049}{0.078} & \er{4.558}{0.019}{0.040} & \er{0.240}{0.019}{0.025} & \er{0.738}{0.013}{0.012} & \er{10.93}{0.83}{1.10} & \er{0.597}{0.047}{0.062} & \er{ 956}{15}{ 9} \\
WASP-69   & \er{0.04555}{0.00161}{0.00130} & \er{0.843}{0.092}{0.070} & \er{4.538}{0.037}{0.035} & \er{0.263}{0.025}{0.022} & \er{1.063}{0.025}{0.017} & \er{ 5.77}{0.49}{0.50} & \er{0.219}{0.019}{0.021} & \er{ 995}{14}{16} \\
HD 189733 & \er{0.02984}{0.00079}{0.00066} & \er{0.719}{0.059}{0.047} & \er{4.531}{0.025}{0.025} & \er{1.037}{0.056}{0.046} & \er{1.154}{0.022}{0.015} & \er{19.29}{0.71}{0.75} & \er{0.674}{0.026}{0.030} & \er{1230}{14}{16} \\
WASP-52   & \er{0.02742}{0.00063}{0.00041} & \er{0.898}{0.064}{0.040} & \er{4.568}{0.011}{0.008} & \er{0.467}{0.028}{0.021} & \er{1.300}{0.030}{0.019} & \er{ 6.85}{0.24}{0.25} & \er{0.212}{0.008}{0.009} & \er{1349}{ 9}{14} \\
HAT-P-26  & \er{0.04342}{0.00309}{0.00314} & \er{0.609}{0.139}{0.123} & \er{4.378}{0.085}{0.098} & \er{0.048}{0.009}{0.009} & \er{0.596}{0.015}{0.012} & \er{ 3.37}{0.61}{0.63} & \er{0.228}{0.041}{0.044} & \er{1081}{41}{35} \\
HD 97658  & \er{0.07916}{0.00367}{0.00276} & \er{0.735}{0.107}{0.074} & \er{4.578}{0.052}{0.046} & \er{0.023}{0.003}{0.003} & \er{0.208}{0.009}{0.007} & \er{13.21}{1.90}{1.83} & \er{2.561}{0.418}{0.401} & \er{ 764}{13}{16} \\
WASP-6    & \er{0.03827}{0.00092}{0.00079} & \er{0.662}{0.049}{0.040} & \er{4.453}{0.020}{0.019} & \er{0.415}{0.023}{0.018} & \er{1.138}{0.025}{0.020} & \er{ 7.96}{0.30}{0.28} & \er{0.282}{0.013}{0.012} & \er{1196}{13}{14} \\
WASP-74   & \er{0.03367}{0.00078}{0.00059} & \er{1.114}{0.079}{0.058} & \er{4.132}{0.018}{0.016} & \er{0.790}{0.038}{0.030} & \er{1.320}{0.028}{0.020} & \er{11.23}{0.32}{0.35} & \er{0.343}{0.011}{0.014} & \er{1944}{19}{24} \\
HAT-P-41  & \er{0.04481}{0.00090}{0.00183} & \er{1.653}{0.102}{0.194} & \er{4.159}{0.021}{0.042} & \er{0.886}{0.102}{0.136} & \er{1.773}{0.035}{0.052} & \er{ 6.99}{0.87}{1.01} & \er{0.159}{0.022}{0.023} & \er{1937}{42}{19} \\
HAT-P-32  & \er{0.03501}{0.00044}{0.00082} & \er{1.238}{0.047}{0.085} & \er{4.341}{0.006}{0.010} & \er{0.899}{0.152}{0.176} & \er{1.833}{0.024}{0.043} & \er{ 6.63}{1.18}{1.22} & \er{0.146}{0.027}{0.027} & \er{1836}{28}{12} \\
WASP-76   & \er{0.03431}{0.00097}{0.00072} & \er{1.644}{0.143}{0.102} & \er{4.133}{0.015}{0.013} & \er{0.998}{0.058}{0.045} & \er{1.903}{0.053}{0.038} & \er{ 6.83}{0.14}{0.15} & \er{0.145}{0.004}{0.005} & \er{2206}{19}{28} \\
WASP-62   & \er{0.05465}{0.00182}{0.00223} & \er{1.118}{0.115}{0.132} & \er{4.301}{0.041}{0.047} & \er{0.531}{0.037}{0.057} & \er{1.338}{0.017}{0.027} & \er{ 7.35}{0.52}{0.70} & \er{0.222}{0.017}{0.021} & \er{1443}{32}{23} \\
WASP-101  & \er{0.05082}{0.00156}{0.00129} & \er{1.362}{0.129}{0.101} & \er{4.347}{0.030}{0.029} & \er{0.502}{0.050}{0.043} & \er{1.369}{0.029}{0.019} & \er{ 6.63}{0.57}{0.57} & \er{0.195}{0.017}{0.017} & \er{1567}{21}{27} \\
WASP-121  & \er{0.02510}{0.00057}{0.00036} & \er{1.298}{0.091}{0.055} & \er{4.236}{0.012}{0.010} & \er{1.124}{0.075}{0.058} & \er{1.753}{0.039}{0.024} & \er{ 9.07}{0.44}{0.43} & \er{0.209}{0.010}{0.011} & \er{2470}{19}{38} \\
WASP-79   & \er{0.05202}{0.00091}{0.00110} & \er{1.399}{0.075}{0.087} & \er{4.195}{0.012}{0.013} & \er{0.842}{0.077}{0.081} & \er{1.625}{0.026}{0.033} & \er{ 7.90}{0.70}{0.69} & \er{0.196}{0.018}{0.017} & \er{1787}{23}{19} \\
KELT-7    & \er{0.04421}{0.00154}{0.00131} & \er{1.541}{0.166}{0.133} & \er{4.153}{0.036}{0.037} & \er{1.294}{0.180}{0.164} & \er{1.502}{0.034}{0.019} & \er{14.21}{1.78}{1.85} & \er{0.382}{0.047}{0.052} & \er{2056}{30}{42} \\
\hline
\end{tabular}
\label{tab:output}
\end{table*}

Based on the arguments above, we collected measurements of $P$, $e$, $i$, $\rho_\star$ and $K_\star$ from the literature, and added them to our new determinations of $T_\star$ and $R_\star$. These were then used to determine the physical properties of the systems via the equations above. The uncertainties in all input parameters were propagated using Monte Carlo techniques.
Table\,\ref{tab:input} contains the values we used for each system, along with references to their source and the discovery paper of each system. In most cases we adopted literature results where possible, but in the case of WASP-29, WASP-76 and WASP-79 we determined new physical properties from more recently available lightcurves to ensure the values were as precise and accurate as possible (see Appendix \ref{sec:new_lightcurves}).
Table\,\ref{tab:output} contains the results.
The mean random uncertainties are 10 per cent in planetary mass, 2 per cent in planetary radius and 9 per cent in stellar mass.

\subsubsection{Exoplanet Parameters - Discussion} % (fold)
\begin{figure}
    \includegraphics[width=\columnwidth]{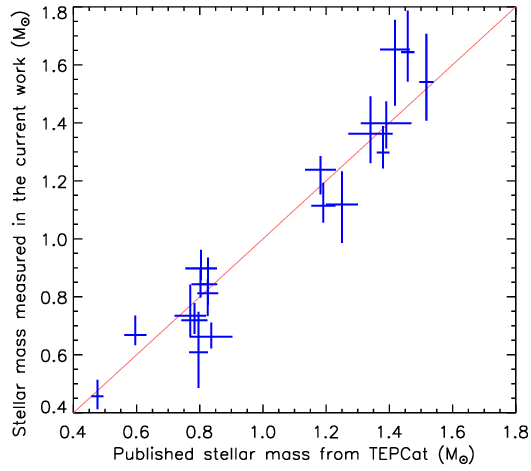} \\
    \caption{\label{fig:M1} Comparison between the host star masses measured in the literature  and our own results. The red line is parity.}
\end{figure}

\label{sec:exo-discuss}
% Comment on the derived exoplanet properties themselves.
\begin{figure}
	\includegraphics[width=\columnwidth]{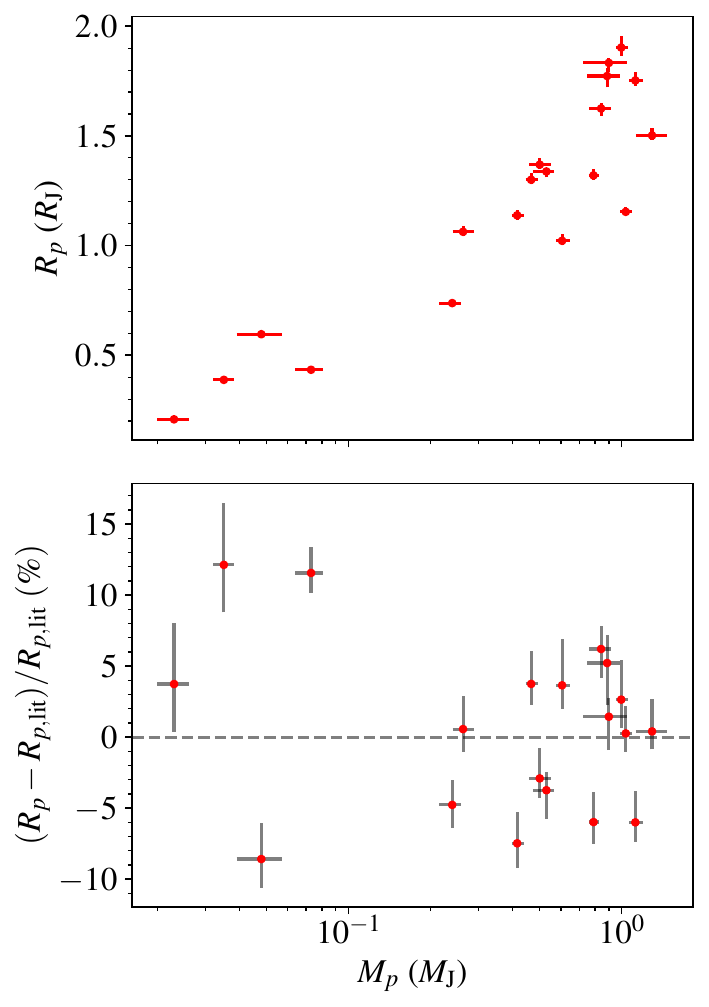}
	\caption{The revised exoplanet radii shown in the $M_{\rm p} - R_{\rm p}$ plane (top), and the difference in per cent between the revised and literature radii provided by TEPCat (bottom). Several revised exoplanet radii agree well with literature determinations. However, many are in disagreement, with the residual exhibiting a RMS difference of 5 {per cent}. }
	\label{fig:planet-mass-radius}
\end{figure}
The precision of the exoplanet measurements presented is significantly improved compared to the literature, with the latter having mean precisions in radius that are about 40 per cent larger than those presented here.
The differences in radius between the literature values and the ones derived in this paper for the exoplanets in our sample, are shown in the lower panel of Fig.\,\ref{fig:planet-mass-radius}.
Again the literature values were taken from TEPCat and mostly correspond to values given in the publications cited in Table\,\ref{tab:input}.  
The mean difference is 0.2 {per cent}, the median 0.6 {per cent}, indicating good correspondence between the methods. 
A quarter of the sample are consistent with previous literature determinations, with the measured radii of WASP-69 b, HD 189733 b, HAT-P-32 b, WASP-121 b and KELT-7 b agreeing to within $1\sigma$. 
The disparity between this and the two thirds agreement in host star luminosities further supports our previous assertion that the uncertainties we presented in section \ref{sec:g_band_test} for the synthetic $G$-band magnitudes of host stars based on the literature parameters are over-estimates.
Fig.\,\ref{fig:M1} shows that there is good agreement between our calculated masses for the host stars with those from the literature.
We also note that our uncertainties in stellar mass and planetary radius are close to those predicted by the simulation of \cite{2018ApJ...862...53S} for a similar method, though our uncertainties in planetary masses are about a factor two larger.

\section{Sources of Systematic Error}
\label{sec:sources-of-systematic-error}
Although we have already tested for some sources of systematic error by comparison with previous work (Sections \ref{sec:test_teff} and \ref{sec:test_rad}), there are some sources of error such tests will not detect.
Hence in this section, we will consider some further causes of systematic uncertainties and provide quantitative estimates of their effects.
Since these sources of systematic error are related to determining the stellar parameters, we will compare with the mean random uncertainties for our sample of 2\% in stellar radius and 50\,K in $T_{\rm SED }$ (section \ref{sec:results}).

\subsection{Extinction}
\label{sec:extinction-discussion} 
In \citet{Morrell:2019aa} we imposed an upper limit of $d < 100$pc to avoid systematics resulting from extinction---at such proximity the effect of interstellar extinction is negligible. 
However, studies of exoplanet hosts require that this constraint be relaxed, necessitating the approach described in section \ref{sec:extinction}.
However, the determination of the extinction will itself have errors, and in this section we explore their effects on our parameter determinations.

\subsubsection{Variation in the ratio of total to selective absorption} 
The source of the largest error due to under-constrained extinction is probably the variation of $R_{55}$ within our galaxy.
\cite{2016ApJ...821...78S} find that $R_V$ has an RMS variation of $\pm0.18$ mags over a set of galactic sightlines.
We therefore folded model atmospheres through filter responses with $R_{55}$ values of $3.02\pm0.18$ for a temperature of 6000\,K and $\log(g)=5.0$.
We found for $A_{55}=0.75$\,mags this change in $R_{55}$ gave a shift of $\pm0.015$\,mags in $G_{\rm BP}-G_{\rm RP}$ which corresponds to temperature shifts of $\pm50$\,K.
Hence $A_{55}=0.75$ is the limit beyond which uncertainties in $R_{55}$ will cause larger uncertainties in $T_{\rm SED}$ than our typical random uncertainty.

The error in extinction will also lead to an error in the luminosity, and hence in the inferred radius.
The largest change in predicted extinction for our range of $R_{55}$ is in the $J$-band and is  only $\pm0.013$\,mags, so even if this were reproduced over all bands the implied change in luminosity would only change the radius by $\pm0.6$ per cent, significantly less than our random error.
So we conclude that the Galactic variation in $A_{55}$ will drive errors in the derived temperatures which become dominant over the random uncertainties for $A_{55}>0.75$.

This analysis can also be used show that systematics in the extinction law will have little impact on our uncertainties.
The Galactic extinction law is characterised by $R$, and the variations in $R$ between mean galactic laws are much smaller than the variation in $R$ tested above, and so will produce smaller errors.

\subsubsection{Errors in the measured extinction}
\label{sec:ext_error_calc}
Before addressing other sources of error due to the extinction measurements we need to determine how a given error in the extinction measurement would translate into an error in the stellar parameters.
We determined this using grids of synthetic photometry where the stars were reddened by $E(44-55) = 0.01\ {\rm and}\ 0.1$.
From these grids we then produced a catalogue of simulated reddened stars by iterating over $T_{\rm eff}$.
For each $T_{\rm eff}$ we determined an appropriate $R_\star$ using the $T_{\rm eff}-R_\star$ relationship prescribed by the 4\,Gyr isochrone from \citet{Baraffe:2015aa}, which remains monotonic and well-defined for $3000{\rm K} \lesssim T_{\rm eff} \lesssim 6000{\rm K}$, hence avoiding problems with interpolation.
The accompanying uncertainty for each synthetic magnitude was derived by Monte Carlo sampling the cumulative distribution function (CDF) of the observed uncertainties for the corresponding photometric band in our input catalogue. 
Each simulated star was then placed at $10$\,pc, and fitted with the unreddened grid of CIFIST models used for the measurements performed in this work. 

\begin{figure}
	\includegraphics[width=\columnwidth]{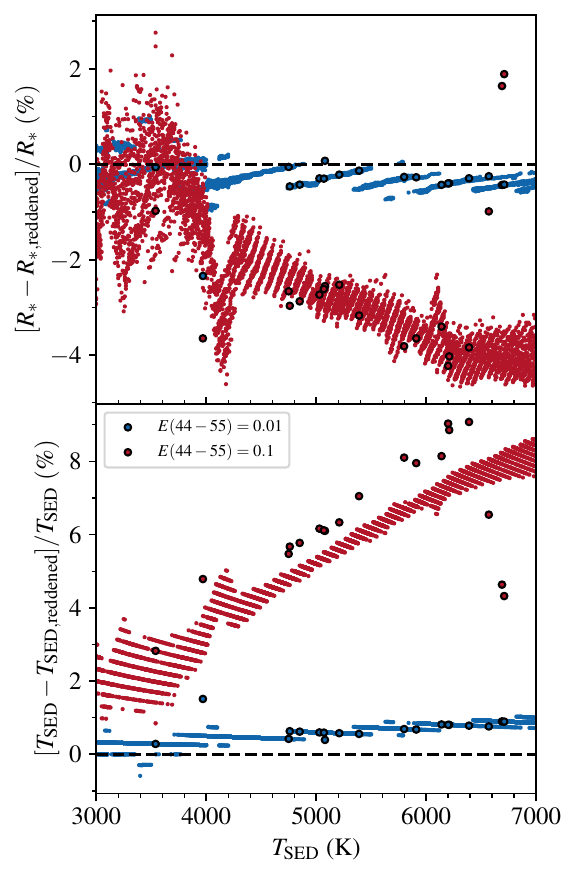}
	\caption{The systematic effect of fitting the SED of reddened synthetic stars with unreddened grids, showing the radius (top) and temperature (bottom) residuals.
    $T_{\rm SED}$ and $R_*$ refer to the temperatures and radii of the synthetic reddened star, and $T_{\rm SED,reddened}$ and $R_{*{,\rm reddened}}$ to the radii and temperatures derived by fitting those stars with unreddened grids.
    The points are coloured blue for synthetic stars $E(44-55) = 0.01$ and red for stars with $E(44-55) = 0.1$.   
    The circles show the effect of fitting the PanCET sample with reddened grids.
    For these the y-axis is the temperature derived by fitting the star with unreddened grids subtracted from the temperature derived from fitting reddened grids, as a percentage of the former.}
	\label{fig:extinction-resid}
\end{figure}

The difference between the parameters of the reddened stars and the results from fitting them with unreddened grids is presented in Fig.\,\ref{fig:extinction-resid}.
This figure shows that any extinction in the sight-line to a planet host star will attenuate and redden the observed stellar SED, hence the fitted measurements moving to cooler $T_{\rm SED}$ at larger $R_\star$.
The colour--dependent nature of extinction means that this systematic correlates well with spectral type -- bluer stars see the most extreme attenuation and reddening, whilst an M-dwarf subjected to an equivalent extinction only sees systematics of $\pm 2$ per cent and $+2$ per cent for $R_\star$ and $T_{\rm SED}$ respectively, due to the majority of their flux already being emitted at IR wavelengths.
Our analysis shows that stars with errors in $E(44-55)$ of $0.01$\,mags (or $A_{55} = 0.03$\,mags) will have systematics that might meet the 1 per cent level for both extinction and radius measurements across the mass range.
Thus it is at this point that they would introduce errors comparable with the random uncertainties in temperature.

\subsubsection{Error due to misplacing the star in the extinction map}
The uncertainty in the parallax of the star will introduce an uncertainty in where in the extinction map the star should be placed.
To assess the effect of this error we considered stars brighter than $G=13$ as these have a roughly constant uncertainty in parallax of 0.025\,mas.
We then placed objects in the extinction map at 1 degree intervals in the Galactic Plane ($|b|<10$) and up to 2000\,pc away in steps of 100\,pc, and calculated the difference in extinction caused by moving the objects by 0.025\,mas in parallax.
At $A_{55}=0.75$, the point at which uncertainties in $R_{55}$ will introduce uncertainties larger than our typical random uncertainty in temperature, only 12 per cent of sightlines have errors in $A_{55}$ due to object placement which exceed the 0.03\, mags the limit shown in section \ref{sec:ext_error_calc} to drive an error larger than our random uncertainties.
Hence the error due to uncertainties in $R_{55}$ exceeds that due to errors in the parallax of the object.

\subsubsection{Error due to uncertainties in the extinction map}
We used the same set of sightlines described above to sample the errors \cite{2022AA...664A.174V} provide for their extinction map.
Proceeding in the same way we find that about 0.05 per cent of sightlines at $A_{55}=0.75$ have uncertainties exceeding 0.03 mags, and so the random uncertainties for the extinction map will always be a much smaller source of uncertainty than the uncertainty in $R_{55}$ for $A_{55}< 0.75$.

\subsubsection{Summary for extinction}
The dominant source of error from the estimated extinction is the variation of $R_{55}$ in the Galaxy, and this is smaller than our random uncertainties for $A_{55}<0.75$.
This is a factor of 2.5 higher than the extinction to any star discussed in this paper, and a factor 20 above the median.
This is not an insurmountable problem if one wished to apply the technique at higher extinctions, as Galactic maps of $R$ are becoming available \citep[e.g.][]{2025Sci...387.1209Z}.
However, the remaining significant known unknown is the systematics in the extinction map, which are hard to assess.
% subsection extinction-discussion (end)

\subsection{Metallicity} % (fold)
\label{sec:metallicity-discussion}
In \citet{Morrell:2019aa} we found that the scatter in the measured radii of M-dwarf stars exhibited a correlation with [Fe/H].
However, we demonstrated that this was caused by fitting the distribution of metallicities in the Solar neighbourhood using only Solar-metallicity models.
We have already corrected the M-dwarf radii presented in \autoref{sec:results} using equation 16 of \citet{Morrell:2019aa}.
However, the generalisation of this methodology necessitates that we investigate the systematic error caused by disparity in metallicity across the entire mass-range occupied by our exoplanet hosts. 

\begin{figure}
	\includegraphics[width=\linewidth]{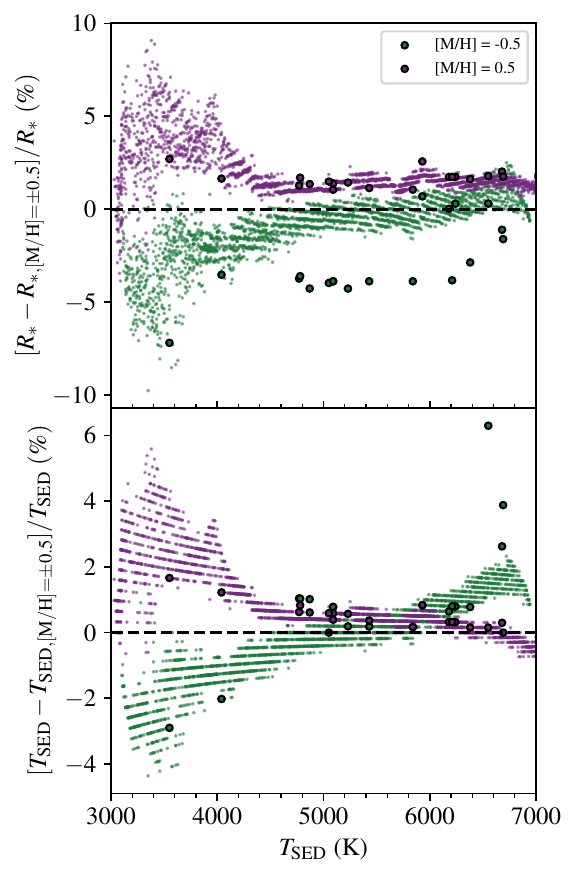}
	\caption{As Fig.\,\ref{fig:extinction-resid}, but illustrating the effects of fitting a supersolar and subsolar synthetic stars with solar metallicity atmosphere grids. 
    $T_{\rm SED,[M/H]=\pm0.5}$ and $R_{*{,\rm [M/H]=\pm0.5}}$ refer to the temperatures and radii  derived by fitting high or low metallicity synthetic stars with solar metallicity grids.
    %The results from fitting the synthetic stars and the PanCET sample are consistent, showing that stars cooler than 4500\ \si{\kelvin} suffer up to a 10\% perturbation in radius if uncorrected. However, stars hotter than 4000\ \si{\kelvin} retrieve the correct radii to within about 1\%. 
    }
	\label{fig:metallicity-resid}
\end{figure}

To investigate different metallicities, we synthesised catalogues at [M/H] = -0.5, 0.0 and +0.5 at Solar abundances.
The atmospheres for this grid are provided by the AGSS2009 stellar atmospheres, which substitute the \citet{Caffau:2011aa} solar abundances used in the CIFIST models for those of \citet{Asplund:2009aa}. 
These grids are limited to a range of $\log(g) = 4.0 - 4.5$, due to this being the range of surface gravity covered by the AGSS2009 atmosphere grids at the required temperatures for all metallicities.
We then fitted these as described for the extinction experiments in Section \ref{sec:ext_error_calc}.
The result of this fitting is shown in Fig.\,\ref{fig:metallicity-resid}. 
As one would intuitively expect, the metallicity-induced discrepancy has a much stronger effect in stars whose $T_{\rm SED} < 4000\ \si{\kelvin}$, where the temperature in the photosphere is cool enough that strong molecular features in the atmosphere become commonplace. 
However, the discrepancy for stars with $T_{\rm SED} > 4000\ \si{\kelvin}$ is at most on the 2.5 per cent level for both $T_{\rm SED}$ and $R_\star$; presumably due to the atmospheres being too hot for said molecular species to form. 
Hence, metallicity systematics should remain at worst comparable to the random uncertainty of the measurements for mid-K to late-F targets ($4500\si{\kelvin} \lesssim T_{\rm SED} \lesssim 6000 \si{\kelvin}$). 
%However, discrepancies of up to $\pm10$ per cent in $R_\star$ and $\pm5$ per cent in $T_{\rm SED}$ can be incurred by not properly accounting for metallicity in low-mass hosts. 
%Equally, there appears to be a systematic of up to $\pm 2$ per cent in measuring $T_{\rm SED}$ for stars earlier than late-F, however there is no equivalent apparent systematic in the measured $R_\star$ in this regime. 
% subsection metallicity-discussion (end)

\subsection{Activity}
\label{sec:activity-discussion}
It is particularly difficult to mitigate against the dynamic, multi-timescale behaviour of starspots and faculae/plages on stellar surfaces. 
It has long been known that convective cells on stellar surfaces are a source of noise for radial velocity observations \citep[e.g.][]{Haywood:2016aa}.
However, \citet{Morris:2020aa} show that the same is true for transit observations, suggesting that stellar activity imposes an uncertainty floor of 3.6 per cent in $R_\star$ for the PLATO mission \citep{Rauer:2014aa,Rauer+25exa}.
As demonstrated in \citet{Morrell:2019aa}, starspots are also able to perturb the temperature and radius measured by our SED fitting methodology. 
However, we also established that, in order to be consistent with a tight M-dwarf MS, filling factors of MS M-dwarfs should typically exhibit a spread of $< 10$ per cent.

To model stars with spots or plage we followed the method presented in Section 4.2.1 of \citet{Morrell:2019aa}. 
The surfaces of the stars were assumed to consist of areas of immaculate photosphere at a temperature $T_{\rm imac}$ and spots or plage at a temperature $T_{\rm spot}$ or $T_{\rm plage}$. 
The fraction of the surface covered by spots or plage is referred to as the filling factor, $\gamma$.
Stars in the 4.3\,Gyr-old cluster M67 can be used to estimate $\gamma$ for starspots as 0.03$\pm$0.04 (RMS) for middle-aged main-sequence G and early K-stars typical of targets for planet-hunting \citep{2022MNRAS.517.2165C}.\footnote{Using the data in their Table 3.}
We further assumed $T_{\rm spot} = 0.8\ T_{\rm imac}$ \citep[see ][]{Berdyugina:2005aa}.
The parameters for plage have thus far only been constrained by observations of the Solar surface, where they have a $\gamma$ of up to about 6 per cent and temperature contrast of $100 - 300\ \si{\kelvin}$, \citep{Oshagh:2014aa,Worden:1998aa,Unruh:1999aa,Meunier:2010aa}.
We adopt the upper end of this temperature difference range and use $T_{\rm plage} = 1.05\ T_{\rm imac}$.

\begin{figure*}
	\includegraphics[width=\textwidth]{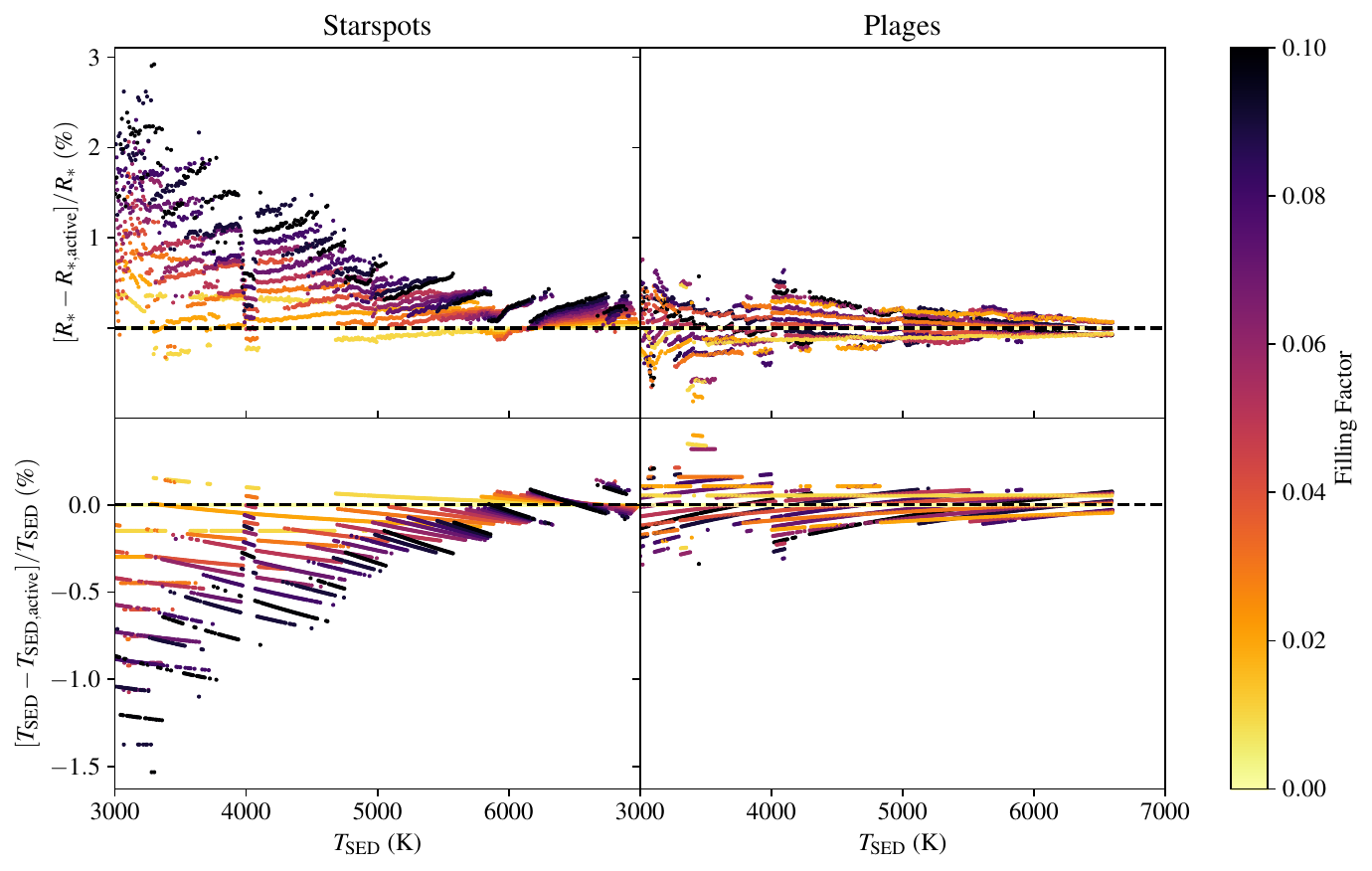}
	\caption{As Fig.\,\ref{fig:extinction-resid}, but illustrating the effects of fitting synthetic stars with starspots (left) and plage (right) with single-temperature grids. 
    $T_{\rm SED,active}$ and $R_{*{\rm,active}}$ refer to the temperatures and radii derived by fitting the synthetic stars with single-temperature grids. 
    %This shows that for reasonable filling factors of up to $\sim 10$ per cent, a scatter of $3$ per cent should be expected in $R_\star$ and about $2$ per cent in $T_{\rm SED}$. This scatter reduces to $< 1$ per cent for temperatures hotter than 5000\ \si{\kelvin}. 
    The plage models truncate at $6600\ \si{\kelvin}$ due to the plage photosphere exceeding the upper $T_{\rm eff}$ limit of the CIFIST model grid. }
	\label{fig:activity-resid}
\end{figure*}

The result of fits to the grids with activity included are shown in Fig.\,\ref{fig:activity-resid}, for filling factors (i.e., fraction of plage or spots) up to and including 0.1. 
From these figures we conclude that our modelled plages have a negligible impact on both $R_\star$ and $T_{\rm SED}$ measurements resulting from our technique; with errors for both remaining below the $1$ per cent level. 
Equally, for K-type stars or earlier, starspots show negligible ($< 1$ per cent) effect on observations for G and early K -stars with filling factors of typical of M67 (0.03), though larger filling factors and/or cooler stars may surpass this limit.
% subsection activity-discussion (end)

\subsection{Photometric zero points and model atmospheres}
\label{sec:zero_and_atmos}
We can address the impact of the uncertainties in the photometric zero points by considering the following thought experiment, which also gives insight into the issues.
Using the SED fitting method to derive the temperature from just two photometric bands is equivalent to determining the temperature from a colour-$T_{\rm eff}$ relationship.
The slope of the colour-$T_{\rm eff}$ relationship then allows us to determine the impact of zero point shifts.
For the comparison we use the BTSettl models for $\log(g)=4.5$.
High temperatures are least sensitive to colour, and so if we take 6\,000\,K the gradient is $-$0.6\,mmag\,K$^{-1}$ in $G_{\rm BP} - K$, which for a 20\,mmag zero point error yields an error in temperature of 33\,K.
This is about half a percent, so given $R_{\star} \propto T^2$ for a fixed luminosity this would be an uncertainty of 1 per cent in radius.
Of course these are certainly over-estimates, as in practice we use many bands which average over zero point uncertainties. 

A similar approach allows us to make an estimate of the uncertainties due to unmodelled effects in the model atmospheres.
If we compare the colours of the Kurucz models \citep{2003IAUS..210P.A20C} with the BTSettl ones, we find a maximum difference of 30\,K at a given colour over the range 3\,500\,K $<$ $T_{\rm eff}$ $<$ 6\,000\,K, again about one percent in radius.
Again this is surely an overestimate, as it is the maximum difference and in addition taking the comparison at face value would imply modern model atmospheres are no better than those of a quarter of a century ago.

\subsection{Missing photometry}
\label{sec:missing_photometry}

A further complication to address involves the relative scarcity of NUV and MIR data, which pin the blue and red end of the SED respectively. 
Given their all-sky nature and wide range of limiting magnitudes, \textit{Gaia} and 2MASS photometry are not of concern.
The situation is particularly severe for the NUV where, for example, only 5 targets of the 19 PanCET sample have suitable NUV photometry. 
Our initial estimate shows that neglecting the WISE or GALEX bands from fitting of the PanCET sample results in differences of $1-2$ per cent in stellar radius. 
In principle this means that using only \textit{Gaia} and 2MASS photometry will have only a small detrimental impact on the determination of exoplanet host radii.

\subsection{Use of UV Photometry}
\label{sec:uv-photometry-discussion}
We have two concerns around the use of UV photometry in SED fitting. 
First is that our current grids of synthetic photometry employ purely photospheric emission, and ignore chromospheric and coronal emission.
It is known that both of these regions exhibit strong line emission; notably H$_\alpha$ and \ce{Ca II} H \& K lines at visible wavelengths, and a number of ionised species in the UV. 
Both H$_\alpha$ and \ce{Ca II} H \& K are weak compared to photospheric emission, and are hence diluted across the wide $G_{\rm BP}$ and $G_{\rm RP}$ bands.
However, the same is not true of NUV, where chromospheric line emission is often much brighter than photospheric emission, especially for late-type stars.
\citet{RodriguezMerino:2005aa} support our concern, finding disagreements in the UV between their photospheric models and spectroscopic observations in Solar-type stars and later, which they attribute to chromospheric emission.
Hence, to confidently fit large samples, we suggest that UV photometry be excluded, or a treatment of coronal and chromospheric emission be included in the synthetic photometry.

\begin{figure*}
    \centering
    \includegraphics[width=\textwidth]{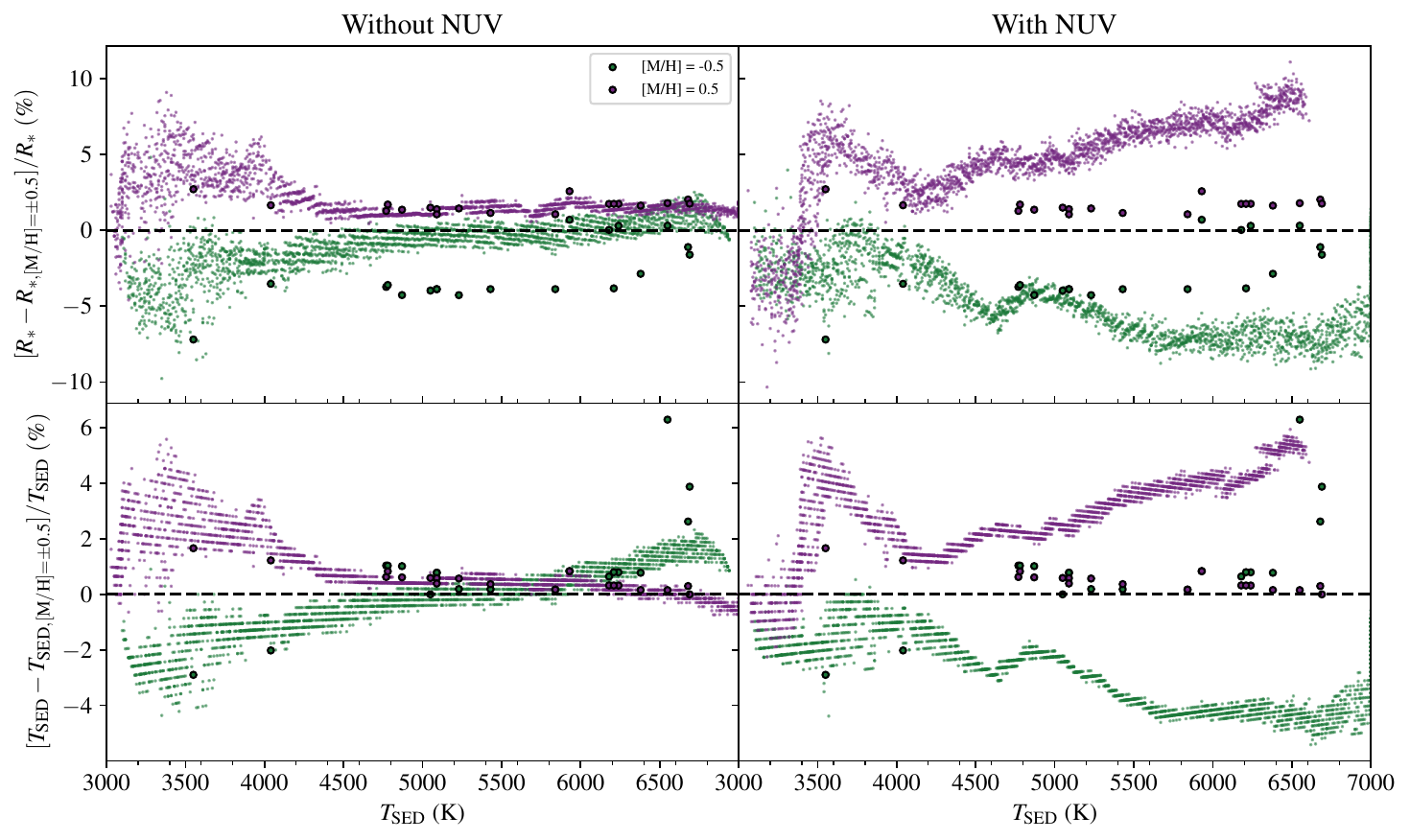}
    \caption{As Fig.\,\ref{fig:metallicity-resid} but showing the predicted systematic due to metallicity when including NUV photometry in the fitting process. We have shown the effect of including the NUV band in this fitting (right) alongside the same data as shown in Fig.\,\ref{fig:metallicity-resid} (left) for comparison. }
    \label{fig:metallicity-resid-uv-comp}
\end{figure*}

\begin{figure*}
    \centering
    \includegraphics[width=\textwidth]{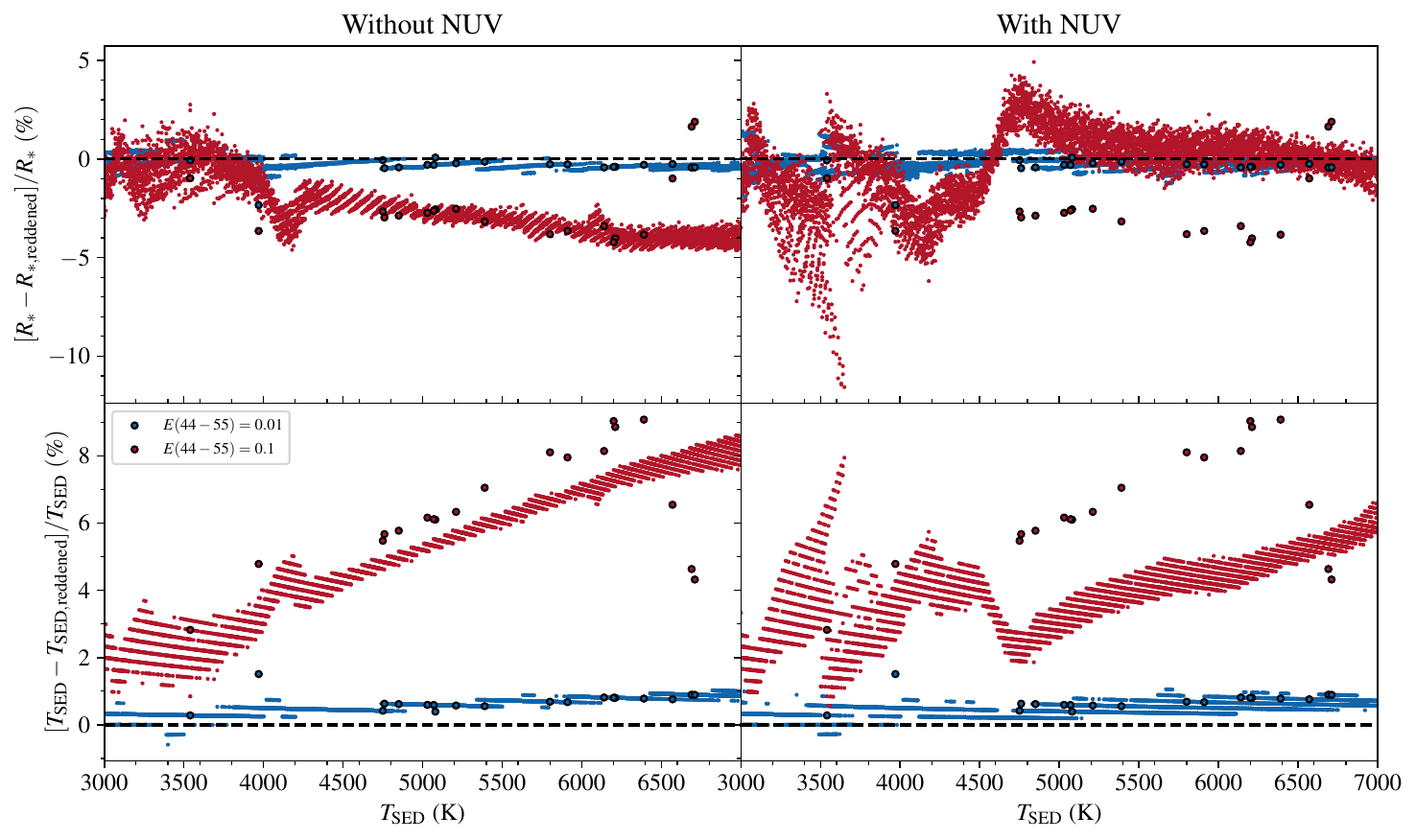}
    \caption{As Fig.\,\ref{fig:extinction-resid} but showing the predicted systematic due to extinction when including NUV photometry in the fitting process. We shown the effect of including the NUV band in this fitting (right) alongside the same data as shown in Fig.\,\ref{fig:extinction-resid} (left) for comparison. }
    \label{fig:extinction-resid-uv-comp}
\end{figure*}

A further concern is how including NUV photometry affects other systematics.
Although we have used observed NUV data for our observational fits, we have not also included this for our simulated catalogues used in \autoref{sec:sources-of-systematic-error}. 
The targets in our sample are hot (early-K and earlier), and have well-studied planets with atmospheres, so are unlikely to exhibit strong activity, or coronal and chromospheric line emission. 
However, including simulated NUV photometry in our model fits affects the observed systematic for metallicity and extinction regardless of activity, as shown in Fig.\,\ref{fig:metallicity-resid-uv-comp} and Fig.\,\ref{fig:extinction-resid-uv-comp} respectively.
The inclusion of NUV accentuates the systematic due to metallicity at the blue end of the SED, bringing the fits of Solar-type stars into parity with their low-mass counterparts, exhibiting a systematic in $R_\star$ and $T_{\rm SED}$ on the order of up to $\pm 10$ per cent and $\pm 6$ per cent respectively, regardless of mass. 
Although including UV photometry does improve the overall residual due to extinction, it does make the extinction systematic less predictable; scattering both the fitted $R_\star$ and $T_{\rm SED}$ back towards the unreddened case with a spread of about $\pm 2$ per cent at an $E(44-55) = 0.1$.

In summary we have strong reservations about the inclusion of UV data in the SEDs.
In addition to the issues explored in this Section, the majority of exoplanet host targets are too red to exhibit strong UV emission, and thus have poor-quality photometry.
Finally we should add the reservations of section \ref{sec:missing_photometry} regarding the scarcity of NUV data and the small impact of omitting it from fits.
All these considerations lead us to recommend that large scale exoplanet characterisation campaigns should not currently employ NUV photometry. 
Note however that for small-scale studies such as this one, where care can be taken over each observational target, supplementing with NUV photometry can be advantageous. 

% subsection uv-photometry-discussion (end)

\subsection{Summary of systematic errors} % (fold)
\label{sec:determining_accurate_radii}
In summarising the effect of systematic errors we begin by addressing those that affect comparisons between parameters for stars that have been determined using the methods outlined in this paper.
The significant errors are due to metallicity, starspots and variation in $R_{55}$.
The relative importance of each of these will depend on the parameters of the stars in question, so as an example we chose the subset of the \cite{2025MNRAS.540.1786F} sample which have metallicities and SED temperatures, as being representative of samples of stars which might be used for radial-velocity planet hunting.
We simulated the effects of metallicity by interpolating (or for |[Fe/H]|$>$0.5 extrapolating) the data of Fig. \ref{fig:metallicity-resid} for each star to find metallicity-driven errors for radius and temperature.  
We then drew a random spot coverage for the star from the Gaussian M67 distribution referred to in Section \ref{sec:activity-discussion} (with values below zero set to zero) and then interpolated the data in Fig. \ref{fig:activity-resid} to derive spot-driven errors.
We found that the effects of metallicity are significantly greater than starspots.
Over the entire sample the sum of both corrections have an RMS of 0.4 per cent in radius and 17\,K in temperature about means of 0.05 per cent and 9\,K respectively.
For comparison the random uncertainties for this sample are 59\,K and 2.0 per cent.
As the largest extinction in the sample is $A_{55}=0.07$, with a median of 0.016\,mags the errors due to $R$ will be negligible compared to the above.
Hence the star-to-star systematic variations are about one third of the random uncertainties for this sample.

The systematic errors which are important when comparing radii and temperatures derived using this SED technique to those from other methods are those due to the photometric zero point errors, and any systematic differences in the temperature scales used. 
Errors in the photometric zero points will manifest themselves as differences in the temperature scales, so we should take the larger of the two errors, rather than combining them.
There are shifts in temperature comparable to the random errors between the SED measurements and the spectroscopic determinations of  \cite{2025MNRAS.540.1786F} (see their Section 5.5.1) and \citet{2016ApJS..225...32B} (section \ref{sec:test_teff}).
Given just these two results, and the fact there is no significant shift between the SED results presented here and the IRFM data of \cite{2010A&A...512A..54C} one might conclude that there there is a difference between the photometric and spectroscopic temperature scales of 30-50\,K.
However, the absence of such a shift between the SED temperatures and those of the Solar twins and analogues (section \ref{sec:test_teff}) argues against this.
What is clear is that there are shifts of several tens of Kelvin between temperature scales, and we note this also corresponds to the maximum likely photometric zero point error of 30\,K.
Hence to give a total systematic uncertainty for comparisons between methods we should probably add 40\,K in quadrature to the 17\,K likely uncertainty for the internal error given above, and simply add the 9\,K shift, to give an error of 52\,K, i.e. similar to the random uncertainty.
In addition we draw attention to the upturn in the offset between spectroscopic and photometric methods seen in the comparison of our SED temperatures and the spectroscopic data of \cite{2025MNRAS.540.1786F} and \citet{2016ApJS..225...32B} for temperatures greater than 6\,000\,K, which suggests a larger systematic for these temperatures. 

Our conclusions are that for comparisons of parameters for different stars determined by the methods outlined in this paper, the random uncertainties are larger than any systematics for our example sample.
However, for comparisons with parameters determined by other techniques, one should allow a systematic of 54\,K, with a corresponding error of 2 per cent in radius.
% subsection determining_accurate_radii (end)

% section discussion (end)

\section{Conclusions} % (fold)
\label{sec:conclusion}

We have presented a revised SED fitting method that can be readily applied to measuring the temperatures and radii of exoplanet host stars (section \ref{sec:star_params}).
\hfil\break\phantom{}\hspace{-1.0mm}  
(i) The mean random uncertainty using this method for the PanCET sample we studied is 50\,K in temperature and 1.9 per cent in stellar radius (section \ref{sec:results}).
\hfil\break\phantom{}\hspace{-1.0mm}
(ii) We have investigated the systematic errors that affect comparisons between stars due to magnetic activity or the model atmospheres we have used, or from incorrect determination of the metallicity or extinction.
For an example sample of FGK stars suitable for radial velocity planet hunting we find these errors lie significantly below the random uncertainties (section \ref{sec:sources-of-systematic-error}).
\hfil\break\phantom{}\hspace{-1.0mm}
(iii) When comparing our results with those from other techniques we find differences in the temperature scales could drive systematic differences comparable with the random uncertainties for temperatures below 6\,000\,K, and larger errors above that temperature (Section \ref{sec:sources-of-systematic-error}).

This method of measuring stellar temperatures and radii has the following advantages over commonly used methods.
\hfil\break\phantom{}\hspace{-1.0mm}
(i) In its simplest form the method uses only archival data (sections \ref{sec:a_new_technique} and \ref{sec:star_params}), avoiding the requirement for new observations.
\hfil\break\phantom{}\hspace{-1.0mm}
(ii) Since the data are all-sky, the measurements of radius and temperature are self-consistent.
\hfil\break\phantom{}\hspace{-1.0mm}
(iii) For the cooler stars with metallicities more than 0.5\,dex from solar the method can be adopted to include a spectroscopic metallicity estimate (section \ref{sec:sources-of-systematic-error}). 

We have discussed how these stellar temperatures and radii can be used to derive parameters for transiting and non-transiting exoplanets (section \ref{sec:application}) and demonstrated this for 19 well-studied exoplanet hosts from the PanCET sample, revising the measurements of the parameters of their planets (section \ref{sec:planet_params}).
Previous determinations for this sample have mean uncertainties in radius which are about 40 per cent larger than those presented here (Section \ref{sec:exo-discuss}), which are a considerable improvement over existing methods. 
We have discussed how, for these transiting systems with radial velocity curves, the method is independent of models of stellar structure and priors for stellar age (Sections \ref{sec:RV+trans} and \ref{sec:planet_params}).

% section conclusion (end)

\section*{Acknowledgements}

We are grateful to an anonymous referee whose constructive scepticism has helped us improve the paper in several respects.
% Staff
SM was in receipt of an Science and Technology Facilities (STFC) Council studentship when much of this work was carried out. 
JS acknowledges support from STFC under grant number ST/Y002563/1.

% Gaia 
This work has made use of data from the European Space Agency (ESA)
mission {\it Gaia} (\url{https://www.cosmos.esa.int/gaia}), processed by
the {\it Gaia} Data Processing and Analysis Consortium (DPAC,
\url{https://www.cosmos.esa.int/web/gaia/dpac/consortium}). Funding
for the DPAC has been provided by national institutions, in particular
the institutions participating in the {\it Gaia} Multilateral Agreement.

%2MASS
This publication makes use of data products from the Two Micron All Sky Survey, which is a joint project of the University of Massachusetts and the Infrared Processing and Analysis Center / California Institute of Technology, funded by the National Aeronautics and Space Administration and the National Science Foundation.

%allWISE
% Source: http://wise2.ipac.caltech.edu/docs/release/allsky/
This publication makes use of data products from the Wide-field Infrared Survey Explorer, which is a joint project of the University of California, Los Angeles, and the Jet Propulsion Laboratory/California Institute of Technology, funded by the National Aeronautics and Space Administration.

\section*{Data availability}
The data underlying this article are available in the VizieR Catalogue service at \href{http://vizier.cds.unistra.fr/}{http://vizier.cds.unistra.fr/}, and can be accessed with `I/355/gaiadr3' (Gaia DR3), `II/246/out' (2MASS), and `II/328/allwise' (AllWISE).
The underlying geometric distances are available in the Astronomisches Rechen-Institut (ARI) {\it Gaia} Services Portal, and can be accessed at \href{https://gaia.ari.uni-heidelberg.de/}{https://gaia.ari.uni-heidelberg.de/}.
The underlying GALEX GR6+/7 data are available in the Barbara A. Mikulski Archive for Space Telescopes, available at \href{https://galex.stsci.edu/GR6/}{https://galex.stsci.edu/GR6/}. 
The TESS data used in this work are available in the Barbara A. Mikulski Archive for Space Telescopes, available at \href{https://mast.stsci.edu/portal/Mashup/Clients/Mast/Portal.html}{https://mast.stsci.edu/portal/Mashup/Clients/Mast/Portal.html}

\bibliographystyle{mnras} 
\bibliography{exoplanet-radii}

\appendix

\section{History}
\label{sec:history}

As we stated in section \ref{sec:a_new_technique}, the method we present here is really a development based on a long history of photometric stellar parameter determination.
In this appendix we aim to show the relationship between the method presented here and similar methods in the literature, at the same time as giving proper recognition to the work that has gone before.

The early history of using flux measurements to determine the angular diameter of stars is briefly reviewed by \cite{1972A&AS....7..257W}.
The key concept is that the square of the angular diameter of a star is proportional to its flux at the Earth in a given band divided by its surface brightness in that band.
The difficulty is that the surface brightness depends on the temperature.
However temperatures can be determined, sometimes through a colour, which leads to the determination of angular radii through surface-brightness colour relationships. Hence a crucial development was the IR flux method \citep[IRFM, e.g. ][]{1979MNRAS.188..847B} which was initially designed to determine angular diameters of stars in a way which depended only weakly on errors in the temperature determination.
The key insight of \cite{1977MNRAS.180..177B} was that as the IR flux originates in the Rayleigh-Jeans tail, it has a much weaker dependence on the temperature than the optical fluxes which are near the peak of the blackbody.
The combination of a narrow-band infrared flux with the bolometric flux yielded the angular diameter.   The method has evolved over the years.
First narrow-band IR measurements were replaced with broadband IR data, modelled by folding model atmospheres through the filter responses \citep[e.g.][]{2004ApJ...609..417R}.
Next broadband optical measurements were used in a similar way, but to estimate the bolometric flux \citep[e.g.][]{ 2005ApJ...626..446R, 2006MNRAS.373...13C}.
As a result the IRFM, though originally conceived as a method of measuring angular diameters which is relatively insensitive to errors in temperature, has now become a benchmark for $T_{\rm eff}$ \citep{1980A&A....82..249B, 2010A&A...512A..54C}.
{\it Gaia} parallaxes have provided the final piece in the puzzle, allowing the angular diameters to be converted into physical stellar diameters.

If we compare the method presented here with the IRFM, we can see that the ingredients are the same.
Both use broadband photometric data (though we add the mid-IR WISE data), and model atmospheres.
The difference lies in the analysis technique; where the IRFM compares IR and bolometric fluxes, we simply carry out a $\chi^2$ fit between the model fluxes folded through the passbands and the photometric data.
We show in section \ref{sec:results} that the difference between temperatures derived using our method and the IRFM is negligible, suggesting the difference in analysis technique is no longer important.

There are, however, other threads in the literature where the flux in various bands are compared to models in a $\chi^2$ sense to determine the temperature (from the shape of the SED) and apparent diameter (from the normalisation).
The earliest example of this approach we have been able to find is \cite{2006A&A...450..735M} whose ``Spectral Energy Distribution Fit'' method uses $V$-band and 2MASS photometry \citep[see also][]{Pecaut:2013aa}.
\cite{2017AJ....153..136S} use a similar technique, but chose to fix the temperature and metallicity of their models from data derived from spectroscopy, leaving just the normalisation as a free parameter.
However, as by this time the {\it Gaia} parallaxes were available, the normalisation provides the luminosity of the star.

A final way of viewing the same fundamental idea is that given {\it Gaia} parallaxes surface-brightness colour relationships can now be used to determine physical rather than angular radii. 
In principle these methods should be less robust than those which use many wavebands, but \cite{2025A&A...699A.325V} show they can produce precise results.

The method presented in this paper is closest to that of \cite{2006A&A...450..735M}, the key difference being that we now have the $Gaia$ parallaxes that allow us to move from apparent radius to physical radius or luminosity \citep[see][for an early example of this]{2017MNRAS.471..770M}.
In addition, though, the $Gaia$ reddening maps mean we can determine a reliable temperature from the SED shape, which negates the requirement for a spectral temperature, and may indeed be more precise (see section \ref{sec:test_teff}).

\section{Analysis of the light curves of WASP-29, WASP-76 and WASP-79}
\label{sec:new_lightcurves}

Several systems in the PanCET sample have been characterised in the past using ground-based light curves of lower quality than what are now available from facilities such as the {\it Kepler} \citep{Borucki16rpph} and Transiting Exoplanet Survey Satellite (TESS) space missions \citep{Ricker+15jatis}. In these cases the newer data allow improved measurements of the light curve parameters and thus a better determination of the physical properties of the system. We therefore analysed three systems, and present the results below. Our analysis included only those TESS data available at the time the work was undertaken (2023 August).

\begin{table*} \centering
\caption{Properties of WASP-29, WASP-76 and WASP-79 measured from their TESS light curves.}
\begin{tabular}{lr@{\,$\pm$\,}lr@{\,$\pm$\,}lr@{\,$\pm$\,}l} \hline
Parameter                   &       \mc{WASP-29}        &       \mc{WASP-76}        &       \mc{WASP-79}                 \\
\hline                      
$P$ (d)                     &   3.9227123  & 0.0000012  &   1.80988113 & 0.00000018 &   3.66239250 & 0.00000045          \\
$T_0$ (BJD$_{\rm TDB}$)     & 58379.95115  & 0.00015    & 59448.895436 & 0.000040   & 58434.866227 & 0.000079            \\
$i$ ($^\circ$)              &        89.32 & 0.68       &        87.29 & 0.50       &        85.60 & 0.13                \\
$r_\star+r_{\rm p}$         &       0.0881 & 0.0021     &       0.2733 & 0.0017     &       0.1547 & 0.0013              \\
$k$                         &      0.09668 & 0.00080    &      0.10738 & 0.00018    &      0.10680 & 0.00021             \\
$r_\star$                   &       0.0803 & 0.0019     &       0.2468 & 0.0015     &       0.1398 & 0.0011              \\
$r_{\rm p}$                 &      0.00777 & 0.00023    &      0.02650 & 0.00020    &      0.01493 & 0.00015             \\
$\rho_\star$ ($\rho_\odot$) &         1.68 & 0.11       &       0.2722 & 0.0057     &       0.3658 & 0.0085              \\
\hline
\end{tabular}
\label{tab:jktebop} 
\end{table*}

\subsection{WASP-29}

\begin{figure}
\includegraphics[width=\columnwidth]{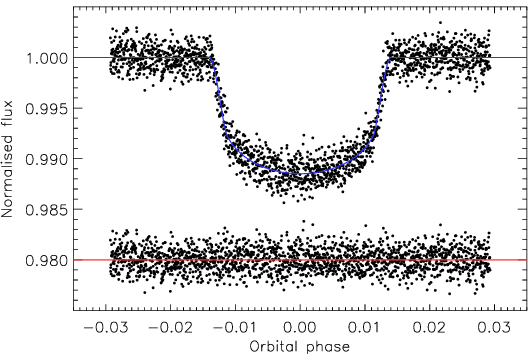} \\
\caption{\label{fig:w29} Best fit of the TESS light curve of WASP-29 (black points) using the
{\sc jktebop} code (red line). The residuals are shown offset to the base of the figure for clarity.}
\end{figure}

\citet{Hellier+10aa} presented the discovery of WASP-29, then the smallest planet discovered by the WASP consortium. Its atmosphere has been found to be featureless via transmission spectroscopy \citep{Gibson+13mn,Wong+22aj} but it has otherwise received relatively little attention. It has been observed by TESS in sectors 2 and 29, in both cases at a cadence of 120\,s. These two sectors contain observations of 11 transits.

We used the {\sc lightkurve} package \citep{Lightkurve18} to download the TESS data in both simple aperture photometry (SAP) and pre-search data conditioning SAP (PDCSAP) versions \citep{Jenkins+16spie}. We rejected points flagged as bad using the ``hardest'' mask option. We found the SAP and PDCSAP data to be very similar so used the PDCSAP data for our analysis. The errorbars were scaled to produce a reduced $\chi^2$ value of unity versus the final fitted model light curve.

Each transit was individually extracted from the light curve alongside another 0.06\,d of data on either side. A straight line was fitted to the out-of-transit data and the transit light curve was normalised to unit flux by dividing by the straight line. The data were then modelled using the {\sc jktebop}\footnote{{\sc jktebop} is written in {\sc fortran77} and the source code is available at {\tt http://www.astro.keele.ac.uk/jkt/codes/jktebop.html}} code \citep{Me13aa}. The fitted parameters were the sum and ratio of the fractional radii ($r_\star+r_{\rm p}$ and $k = {r_{\rm p}}/{r_\star}$), $i$, $P$, a reference time of mid-transit ($T_0$), and the out-of-transit brightness. Limb darkening was included using the power-2 law \citep{Hestroffer97aa,Me23obs2} with the linear coefficient fitted and the nonlinear coefficient fixed to a theoretical value from \citet{ClaretSouthworth22aa}. 

The fitted parameters are given in Table\,\ref{tab:jktebop}, and the best fit to the data is shown in Fig.\,\ref{fig:w29}. The uncertainties were calculated using Monte Carlo simulations \citep{Me08mn}. We calculated $\rho_\star$ from $r_\star$ and $P$ using Equation \ref{eq:rho1}. Our results are significantly more reliable than previous work due to the availability of the space-based light curves.

\subsection{WASP-76}

\begin{figure}
\includegraphics[width=\columnwidth]{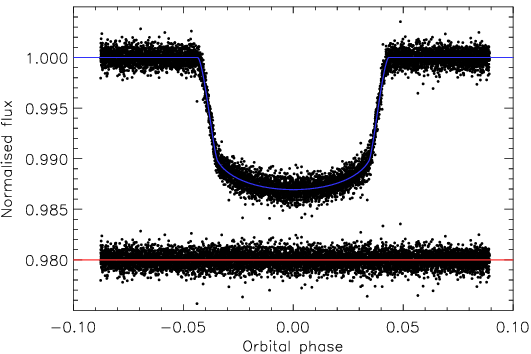} \\
\caption{\label{fig:w76} Best fit of the TESS light curve of WASP-76 (black points) using the
{\sc jktebop} code (red line). The residuals are shown offset to the base of the figure for clarity.}
\end{figure}

WASP-76 was discovered by \citet{West+16aa} and contains a very hot giant planet orbiting an F-star. The system has since been observed extensively via the method of transmission spectroscopy to characterise the atmosphere of the planet \citep[e.g.][]{Kesseli+22aj}.

TESS observed WASP-76 in sectors 30, 42 and 43, in all cases at a cadence of 120\,s. A total of 33 transits were fully observed and included in our analysis. We extracted them from the full light curves and fitted them using {\sc jktebop} in the same way as above. We included the time of mid-transit from \citet{West+16aa} to better constrain the orbital period. 

We had to make one change to the analysis process versus that for WASP-29, concerning the presence of a faint nearby star. \citet{Bohn+20aa} used the VLT/SPHERE instrument to perform $H$-band high-resolution imaging of the WASP-76 system, finding a companion at a distance of 0.436$^{\prime\prime}$ which is fainter than the planet host star by $\Delta H = 2.30 \pm 0.05$\,mag. In an accompanying paper, \citet{Me+20aa} presented new transit photometry and modelled it whilst accounting for this companion. We followed the same procedure in this case but with the now-available TESS data: the brightness ratio of the two stars was propagated from the $H$ band to the TESS passband using {\sc phoenix} theoretical spectra \citep{Allard++12rspta} and included as a third-light contribution in the {\sc jktebop} model \citep{Me10mn}. For reference, we found this contribution to be $0.0062 \pm 0.0009$ in the TESS passband. 

The fitted parameters are given in Table\,\ref{tab:jktebop}, and the best fit to the data is shown in Fig.\,\ref{fig:w76}. The uncertainties were calculated in the same way as for WASP-29.

\subsection{WASP-79}

\begin{figure}
\includegraphics[width=\columnwidth]{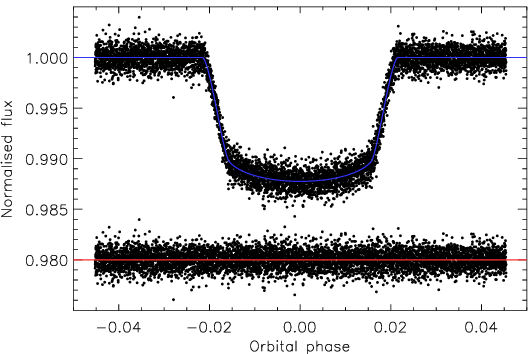} \\
\caption{\label{fig:w79} Best fit of the TESS light curve of WASP-79 (black points) using the {\sc jktebop} code (red line). The residuals are shown offset to the base of the figure for clarity.}
\end{figure}

The discovery of WASP-79 was announced by \citet{Smalley+12aa} based on very limited photometric data. Two complete transits and an improved set of physical properties were presented by \citet{Brown+17mn}. TESS observations were obtained in four sectors (4, 5, 31 and 32) %, with no more currently planned\footnote{\texttt{https://heasarc.gsfc.nasa.gov/cgi-bin/tess/webtess
% \newline
% /wtv.py?Entry=wasp+79}}, 
and cover 26 complete transits with observations taken at a cadence of 120\,s.

Our analysis proceeded identically to that for WASP-29 with the exception that we included the time of mid-transit from \citet{Brown+17mn} to better constrain the orbital period, although in reality this has very little effect on any of the fitted parameters. The fitted parameters are given in Table\,\ref{tab:jktebop}, and the best fit to the data is shown in Fig.\,\ref{fig:w79}. Our results are significantly more reliable than previous work due to the profusion of new transits.

\label{lastpage}
\end{document}